\documentclass[a4paper,11pt]{article}

\usepackage{jheppub} 
\usepackage[normalem]{ulem}  
\usepackage{subcaption}
\usepackage{diagbox}
\usepackage{cleveref}
\usepackage{braket}
\usepackage{bbold}
\usepackage{ulem}

\newcommand{\nn}{\nonumber}
\newcommand{\del}{\partial}

\newcommand{\pa}[1]{\left(#1\right)}

\newcommand{\wilde}{\widetilde}

\graphicspath{{figure/}}

\title{
Perturbative unitarity in quasi-single field inflation
} 

\author[a]{Suro Kim,}

\author[a]{Toshifumi Noumi,}

\author[a]{Keito Takeuchi,}

\author[b]{and Siyi Zhou}

\affiliation[a]{Department of Physics, Kobe University, Kobe 657-8501, Japan}
\affiliation[b]{The Oskar Klein Centre for Cosmoparticle Physics \& Department of Physics, Stockholm University, AlbaNova, 106 91 Stockholm, Sweden}

\emailAdd{s-kim@stu.kobe-u.ac.jp}

\emailAdd{tnoumi@phys.sci.kobe-u.ac.jp}

\emailAdd{tkuc0628@icloud.com}

\emailAdd{siyi.zhou@fysik.su.se}

\preprint{KOBE-COSMO-21-01}

\abstract{
We study implications of perturbative unitarity for quasi-single field inflation with the inflaton and one massive scalar. Analyzing high energy scattering, we show that non-Gaussianities with $|f_{\rm NL}|\gtrsim1$ cannot be realized without turning on interactions which violate unitarity at a high energy scale. Then, we provide a relation between $f_{\rm NL}$ and the scale of new physics that is required for UV completion. In particular we find that for the Hubble scale $H\gtrsim 6\times 10^{9}$ GeV, Planck suppressed operators can easily generate too large non-Gaussanities and so it is hard to realize successful quasi-single field inflation without introducing a mechanism to suppress quantum gravity corrections. Also we generalize the analysis to the regime where the isocurvature mode is heavy and the inflationary dynamics is captured by the inflaton effective theory. Requiring perturbative unitarity of the two-scalar UV models with the inflaton and one heavy scalar, we clarify the parameter space of the $P(X,\phi)$ model which is UV completable by a single heavy scalar.
}

\begin{document} 
\setcounter{tocdepth}{2}
\maketitle
\flushbottom

\section{Introduction}

Perturbative unitarity plays an important role in discovering new physics. Historically, the Higgs boson was predicted by studying unitarity of high-energy scattering of weak bosons. In particular, its mass scale is estimated by the unitarity violation scale in a theory without the Higgs sector~\cite{Lee:1977eg,Lee:1977yc,Dicus:1992vj,Chanowitz:1985hj}. Indeed, the Higgs boson with the mass $125$ GeV was discovered, which completed the Standard Model as a UV complete theory of particle physics~\cite{Chatrchyan:2013lba,Chatrchyan:2012ufa,Aad:2012tfa}.

\medskip
The purpose of this paper is to apply a similar idea in the study of primordial non-Gaussianities~\cite{Maldacena:2002vr}\footnote{
See, e.g., Refs.~\cite{Lerner:2010mq,Giudice:2010ka,Atkins:2010yg,Calmet:2013hia,Barbon:2015fla,Fumagalli:2017cdo,Lee:2018esk,Ema:2020zvg} 
for application of perturbative unitarity in the inflation context, especially in the context of Higgs inflation. Perturbative unitarity was used there to constrain the background dynamics of inflation, whereas the present paper discusses implications for primordial non-Gaussianities.},
in light of recent studies of perturbative unitarity in the Standard Model Effective Field Theory (SMEFT)~\cite{Weinberg:1979sa,Buchmuller:1985jz,Grzadkowski:2010es,Grinstein:2007iv} (see Refs.~\cite{deFlorian:2016spz,Brivio:2017vri} for review articles).
In SMEFT, effects of new physics beyond the Standard Model are encoded into low-energy effective interactions. For example, the Higgs potential in the Standard Model has two parameters characterizing the Higgs vacuum expectation value (vev) and the Higgs mass. In particular, the Higgs trilinear coupling is no more free parameter once the Higgs vev and the Higgs mass are specified. Then, a deviation from the Standard Model value implies a new physics at a higher energy scale, which is specified by studying perturbative unitarity~\cite{Chang:2019vez}. Another typical example is derivative interactions in the scalar sector, i.e., the Higgs boson and the Nambu-Goldstone bosons that are eaten by the gauge bosons. If we focus on two-derivative interactions, perturbative unitarity requires flatness of the field space in the UV complete theory, which in turn specifies the scale of new physics to be the curvature scale of the SMEFT field space~\cite{Nagai:2019tgi}. In this way, recent studies in the SMEFT context have clarified implications of perturbative unitarity for scalar field theories in particular.

\medskip
In this paper, following these developments, we address two issues in the inflation context. The first is about primordial non-Gaussianities in quasi-single field inflation~\cite{Chen:2009zp}: To maintain the flatness of the inflaton potential, it is natural to assume an approximate shift symmetry of the inflaton field. A simplest realization of such a symmetry is to embed  the inflaton field as a phase direction of a complex scalar field, analogously to the Nambu-Goldstone boson in the Higgs potential. If the radial direction has a mass near the Hubble scale, the dynamics of primordial fluctuations is governed by the inflaton fluctuations and the isocurvature modes with the Hubble scale mass. Then, the model is called quasi-single field inflation. In particular, primordial non-Gaussianities in this model accommodate interesting features which can be used to identify the particle spectra during inflation and so they have been studied intensively in the context of ``Cosmological Collider Program"~\cite{Chen:2009zp,Baumann:2011nk,Noumi:2012vr,Arkani-Hamed:2015bza} (see also \cite{Chen:2009we,Assassi:2012zq,Sefusatti:2012ye,Norena:2012yi,Emami:2013lma,Liu:2015tza,Dimastrogiovanni:2015pla,Schmidt:2015xka,Chen:2015lza,Bonga:2015urq,Delacretaz:2015edn,Flauger:2016idt,Lee:2016vti,Delacretaz:2016nhw,Meerburg:2016zdz,Chen:2016uwp,Chen:2016hrz,An:2017hlx,Tong:2017iat,Iyer:2017qzw,An:2017rwo,Kumar:2017ecc,Riquelme:2017bxt,Saito:2018omt,Cabass:2018roz,Dimastrogiovanni:2018uqy,Bordin:2018pca,Arkani-Hamed:2018kmz,Kumar:2018jxz,Goon:2018fyu,Wu:2018lmx,Chua:2018dqh,Wang:2018tbf,McAneny:2019epy,Li:2019ves,Kim:2019wjo,Sleight:2019mgd,Biagetti:2019bnp,Sleight:2019hfp,Welling:2019bib,Alexander:2019vtb,Lu:2019tjj,Hook:2019zxa,Hook:2019vcn,ScheihingHitschfeld:2019tzr,Baumann:2019oyu,Wang:2019gbi,Liu:2019fag,Wang:2019gok,Wang:2020uic,Li:2020xwr,Baumann:2020dch,Kogai:2020vzz,Aoki:2020zbj,Maru:2021ezc}). In this paper, we discuss implications of perturbative unitarity for primordial non-Gaussianities in quasi-single field inflation with the inflaton and one massive scalar (which we call the two-field quasi-single field inflation in the following) and demonstrate that observable non-Gaussianities with the nonlinearity parameter $|f_{\rm NL}|\gtrsim1$ cannot be realized without turning on interactions which violate unitarity at a high energy scale $\Lambda$. We also provide a relation between the scale $\Lambda$ of new physics and the nonlinearity parameter $f_{\rm NL}$.

\medskip
On the other hand, if the isocurvature modes are heavier than the Hubble scale, an effective theory of the inflaton is obtained by integrating out the heavy modes. A useful template for such effective theories of single-field inflation is the so-called $P(X,\phi)$ model, whose Lagrangian contains an arbitrary function of the inflaton $\phi$ and its kinetic operator $X=-\frac{1}{2}(\partial_\mu\phi)^2$ on top of the Einstein-Hilbert term. In particular, higher derivative interactions can source observable non-Gaussianities~\cite{Chen:2006nt}. Since the function $P(X,\phi)$ contains information of the UV theory such as masses, spins, and number of heavy fields, it is of great interests to clarify which parameter space can be realized by which types of UV completion. By doing so, we can draw a ``country map" of the EFT landscape\footnote{
Following recent developments in the swampland program~\cite{Brennan:2017rbf,Palti:2019pca}, our understanding on the consistent EFT parameter space (the EFT landscape) has been increased considerably. The next step useful for phenomenology would be to separate the EFT landscape into several regions (countries) based on detailed properties of the UV theory beyond the universal consistency conditions such as unitarity and causality.}. Based on this motivation, we use perturbative unitarity to clarify the parameter space scanned by two-scalar UV models, i.e., models with the inflaton and one heavy scalar\footnote{
Perturbative unitarity in the EFT of inflation was discussed in Refs.~\cite{Baumann:2011su,Baumann:2014cja,Koehn:2015vvy,Baumann:2015nta,deRham:2017aoj,Fumagalli:2020ody}, where the strong coupling scale within the EFT framework was estimated for example.
In contrast,
the present paper
studies perturbative unitarity constraints
on the UV theory and the corresponding EFT of inflation realized at IR.
}. We demonstrate that cubic and higher order terms in $X$ cannot be generated at IR unless we introduce new physics beyond the two-scalar UV models. We also provide a relation between the size of the $X^3$ coupling and the scale $\Lambda$ of new physics beyond the two-scalar UV models.

\medskip
The organization of the paper is as follows. In Sec.~\ref{QSI}, we review quasi-single field inflation~\cite{Chen:2009zp} with an emphasis on typical energy scales of the model and their relations to primordial non-Gaussianities. In Sec.~\ref{Perturbative}, we study the requirement of perturbative unitarity on quasi-single field inflation and discuss its implication for non-Gaussianities. In particular, we demonstrate that observable non-Gaussianities with $|f_{\rm NL}|\gtrsim 1$ cannot be realized without turning on interactions which require UV completion beyond quasi-single field inflation. In Sec.~\ref{heavymass}, we clarify the parameter space of the $P(X,\phi)$ model scanned by two-scalar UV models that contain the inflaton and one heavy scalar. We conclude in Sec.~\ref{prospects} with discussion of our results.

\section{Quasi-single field inflation}
\label{QSI}

In this section we review quasi-single field inflation~\cite{Chen:2009zp} with an emphasis on typical energy scales of the model and their relations to primordial non-Gaussianities. After summarizing phenomenological aspects of the model,
we demonstrate that in the two-field quasi-single field inflation, the nonlinearity parameter $f_{\rm NL}$ is too small to observe if its main source is renormalizable interactions. This motivates us to study implications of perturbative unitarity in the next section.

\subsection{Phenomenological aspects}
\label{subsec:pheno}
The model consists of two real scalars, $\phi_1$ and $\phi_2$, with a flat field space metric:
\begin{align}
	S = \int d^4x\sqrt{-g}\left[
	\frac{M_{\rm Pl}^2}{2}R
	- \sum_{i=1,2}\frac{1}{2}\partial_\mu \phi_i \partial^\mu \phi_i - \mathcal{V}(\phi_i)
	\right]\,.
	\label{qsi_phiaction}
\end{align} 
The potential, $\mathcal{V}(\phi_i)$, has an approximate $U(1)$ symmetry under the phase shift of $\phi_1+i\phi_2$, which guarantees the flatness of the potential along the angular direction in the field space. More explicitly, in the polar coordinates,
\begin{align}
\phi_1+i\phi_2=re^{i\theta}\,,
\end{align}
the action takes the form,
\begin{align}
S&=\int d^4 x\sqrt{-g}\left[\frac{M_{\rm Pl}^2}{2}R-\frac12r^2 \del_\mu\theta\del^\mu\theta - \frac12  \del_\mu r\del^\mu r - V_r(r)- V_{\rm soft}(\theta)\right]\,.
\end{align}
Here and in what follows we use $\mathcal V$ for the potential in Cartesian coordinates, $(\phi_1,\phi_2)$, and $V$ for that in the polar coordinates, $(r,\theta)$.
$V_r(r)$ is a $U(1)$ symmetric potential invariant under the constant shift of $\theta$ and $V_{\rm soft}(\theta)$ breaks the shift symmetry softly. We assume that $V_r(r)$ has a minimum at $r=r_{\rm min}\neq0$ away from the origin and the symmetry breaking potential, $V_{\rm soft}(\theta)$, is flat enough for the angular coordinate, $\theta$, to play the role of inflaton.

\paragraph{Inflationary backgrounds.}

Consider an inflationary background,
\begin{align}
\label{background}
\theta=\theta_0(t)\,,
\quad
r=r_0(t)\,,
\quad
ds^2=-dt^2+a(t)^2d\boldsymbol{x}^2\,.
\end{align}
The background equations of motion for $\theta$ and $r$ read
\begin{align}
r_0^2
\ddot{\theta}_0+3Hr_0^2\dot{\theta}_0+2r_0\dot{r}_0\dot{\theta}_0
+V'_{\rm soft}(\theta_0)&=0\,,
\\
\ddot{r}_0+3H\dot{r}_0+V'_r(r_0)
-r_0\dot{\theta}_0^2&=0\,,
\end{align}
where $H=\dot{a}/a$ is the Hubble parameter. The Friedmann equations are
\begin{align}
3M_{\rm Pl}^2H^2&=\frac{1}{2}r_0^2\dot{\theta}_0^2+\frac{1}{2}\dot{r}_0^2+V_r(r_0)+V_{\rm soft}(\theta_0)\,,
\\
-2M_{\rm Pl}^2\dot{H}&=r_0^2\dot{\theta}_0^2+\dot{r}_0^2\,.
\end{align}
In the following we use the slow-roll approximation under which $\dot{\theta}_0$ and $r_0$ are approximately constant. Then, the equation of motion for $r$ reduces to
\begin{align}
\label{eom_r}
r_0\dot{\theta}_0^2=V_r'(r_0)\,,
\end{align}
which specifies how much the background trajectory, $r=r_0$,  deviates from the minimum, $r=r_{\rm min}$, of the potential, $V_r(r)$,  due to the ``centrifugal force."
Also, the soft breaking of the $U(1)$ symmetry of $\mathcal{V}(\phi_i)$, i.e., the shift symmetry of $\theta$,
is quantified by the slow-roll parameters as
\begin{align}
\varepsilon&=-\frac{\dot{H}}{H^2}\simeq\frac{M^2_{\rm pl}}{2r_0^2}\left(\frac{V^\prime_{\rm soft}}{V_r+V_{\rm soft}}\right)^2\,,
\\
\eta&=\frac{\dot{\varepsilon}}{H\varepsilon}\simeq-2\frac{M^2_{\rm pl}}{r_0^2}\frac{V^{\prime\prime}_{\rm soft}}{V_r+V_{\rm soft}}+2\frac{M^2_{\rm pl}}{r_0^2}\left(\frac{V^\prime_{\rm soft}}{V_r+V_{\rm soft}}\right)^2\,,
\end{align}
which are assumed to be small, $\varepsilon,|\eta|\ll 1$, during inflation.

\paragraph{Primordial fluctuations.}

Next let us introduce fluctuations around the inflationary background \eqref{background} as
\begin{align}
\theta=\theta_0+\delta\theta\,,
\quad
r=r_0+\sigma\,.
\label{eq:fluctuations}
\end{align}
Under the slow-roll approximation, the action of matter fluctuations in the spatially flat gauge reads
\begin{align}
\nonumber
S&=\int dt d^3xa^3\bigg[
-\frac12r_0^2 \,\del_\mu\delta\theta\,\del^\mu\delta\theta - \frac12  \del_\mu\sigma\del^\mu\sigma - V(\sigma)
\\
\label{action_delta_theta}
&\qquad\qquad\qquad\,\,\,
-r_0 \sigma\left[-2\dot{\theta}_0\dot{\delta\theta}+(\del_\mu\delta\theta)^2\right]
-\frac{1}{2}\sigma^2\left[-2\dot{\theta}_0\dot{\delta\theta}+(\del_\mu\delta\theta)^2\right]\bigg]\,,
\end{align} 
where we neglected metric fluctuations because their contributions to scalar spectra are subdominant. Also we introduced an effective potential,
\begin{align}
V(\sigma)=V_r(r_0+\sigma) -V_r(r_0) -\left[
\frac{\dot{\theta}_0^2}{2}(r_0+\sigma)^2-\frac{\dot{\theta}_0^2}{2}r_0^2\right] ~, 
\end{align}
which has no term linear in $\sigma$ because of the equation of motion~\eqref{eom_r}.

\medskip
It is convenient to introduce a canonically normalized field, $\varphi=r_0\,\delta\theta$, in terms of which the action~\eqref{action_delta_theta} may be reformulated as
\begin{align}
\nonumber
S&=\int dt d^3xa^3\bigg[
-\frac12 \del_\mu\varphi\del^\mu\varphi - \frac12  \del_\mu\sigma\del^\mu\sigma - V(\sigma)
\\
\label{action_varphi}
&\qquad\qquad\qquad\,\,\,
+2\dot{\theta}_0\dot{\varphi}\sigma+\frac{\dot{\theta}_0}{r_0}\dot{\varphi}\sigma^2
-\frac{\sigma}{r_0}(\partial_\mu\varphi)^2
-\frac{1}{2}\frac{\sigma^2}{r_0^2}(\partial_\mu\varphi)^2
\bigg]\,,
\end{align} 
where the first two terms in the second line are proportional to $\dot{\theta}_0$ and break the Lorentz symmetry. In particular, the Lorentz symmetry breaking sector accommodates a linear mixing between $\varphi$ and $\sigma$. For later use, we parameterize the effective potential $V(\sigma)$, as
\begin{align}
\label{V_sigma}
V(\sigma)=m^2r_0^2\left[
\frac{1}{2}\left(\frac{\sigma}{r_0}\right)^2+\frac{\lambda_3}{2}\left(\frac{\sigma}{r_0}\right)^3
+\frac{\lambda_4}{8}\left(\frac{\sigma}{r_0}\right)^4
+\mathcal{O}(\sigma^5)
\right]\,,
\end{align}
where we introduced dimensionless parameters $\lambda_i$.

\paragraph{Energy scales.}

The model~\eqref{action_varphi} contains (at least) three energy scales characterized by $m$, $\dot{\theta}_0$, and $r_0$. In this paper except for Sec.~\ref{heavymass}, we consider the following typical parameter regime of quasi-single field inflation:
\begin{enumerate}
\item Isocurvature mass $m$.

First, we assume that the mass of the isocurvature mode, $\sigma$, is of the Hubble scale order: $m\sim H$. This is why the model is called {\it quasi-single field} in comparison to multi-field and effective single-field after integrating out heavy fields.

\item Mixing scale $\dot{\theta}_0$.

As a consequence of the turning background trajectory, there appears a linear mixing between the adiabatic mode $\varphi$, and the isocurvature mode $\sigma$. We assume that the mixing is in a perturbative regime during inflation. In other words we assume that the mixing scale is well below the Hubble scale $\dot{\theta}_0\lesssim  H$, which means that the turning angle per $e$-fold is small enough. Our typical benchmark point will be $\dot{\theta}_0=0.1\times H$.

\item Decay constant $r_0$ of the isocurvature $\sigma$.

The radius, $r_0$, of the turning trajectory characterizes the scale of higher derivative interactions, $\sigma(\partial_\mu\varphi)^2$ and $\sigma^2(\partial_\mu\varphi)^2$. For later convenience, in Eq.~\eqref{V_sigma},  we used the same energy scale as the decay constant of $\sigma$, even though there are no reasons to assume $\lambda_{3,4}=\mathcal{O}(1)$ at least at this moment.

\end{enumerate}

\paragraph{Primordial spectra.}

\begin{figure}[t]
	\centering
	\includegraphics[height=4.5cm]{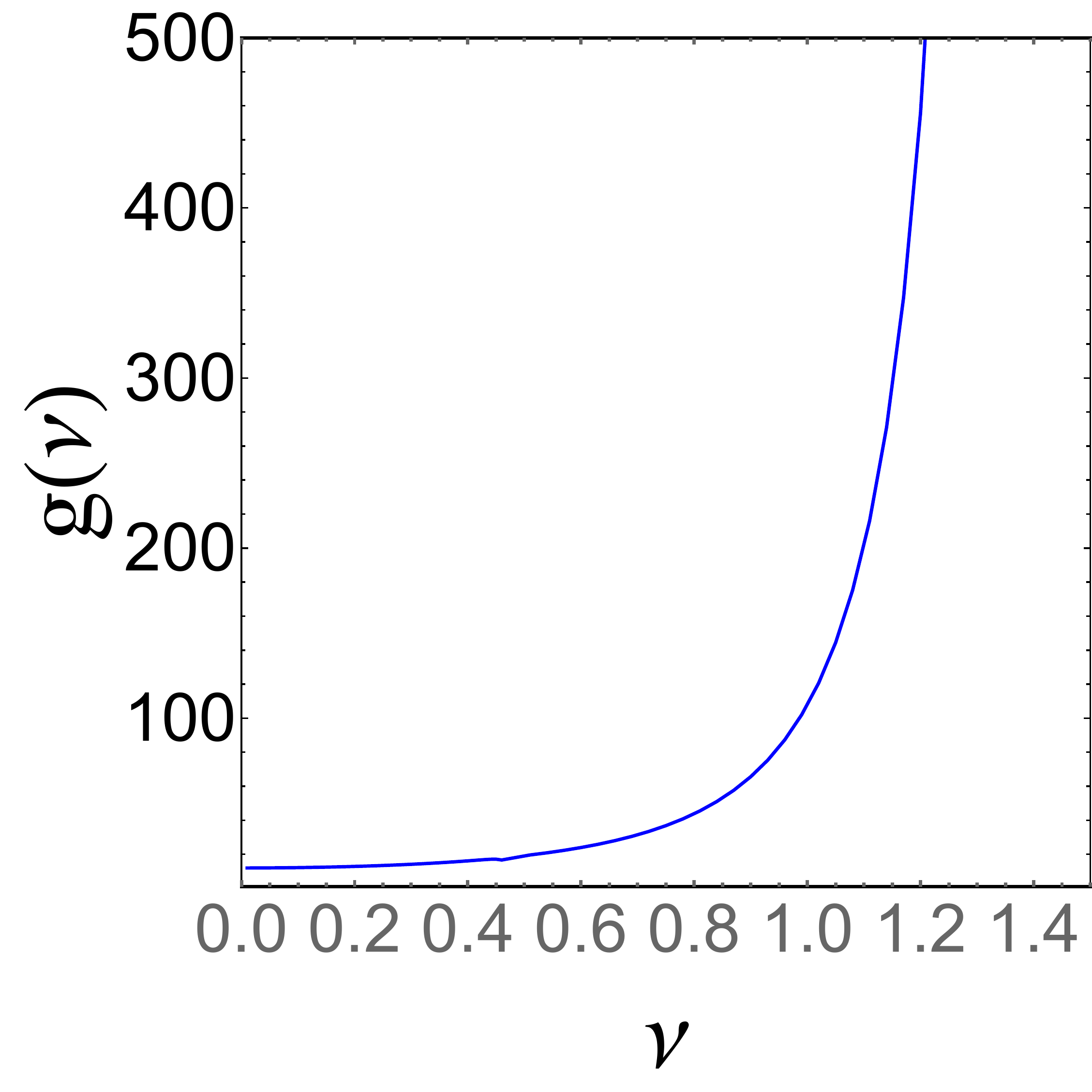} \,
	\includegraphics[height=4.5cm]{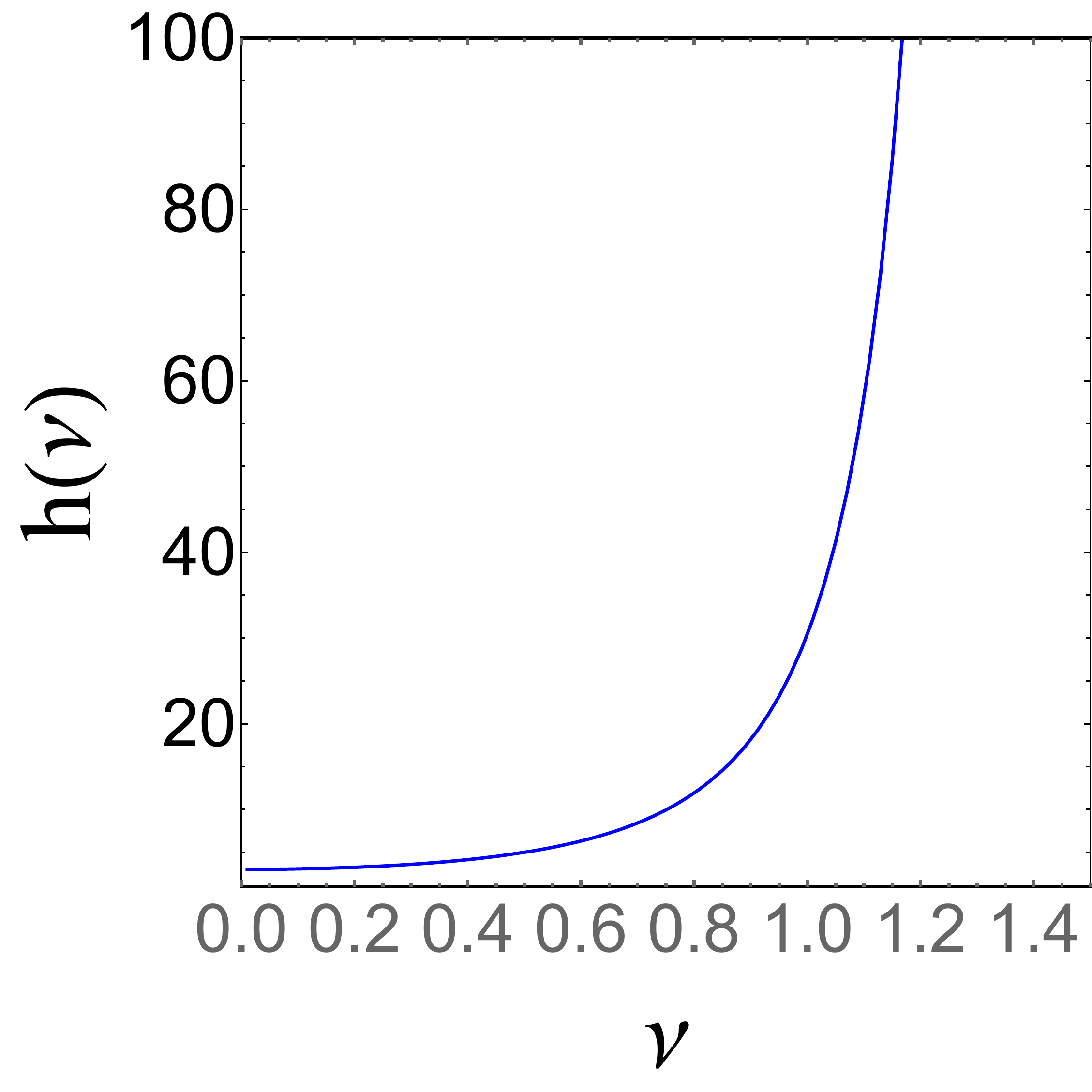}  \,
	\includegraphics[height=4.5cm]{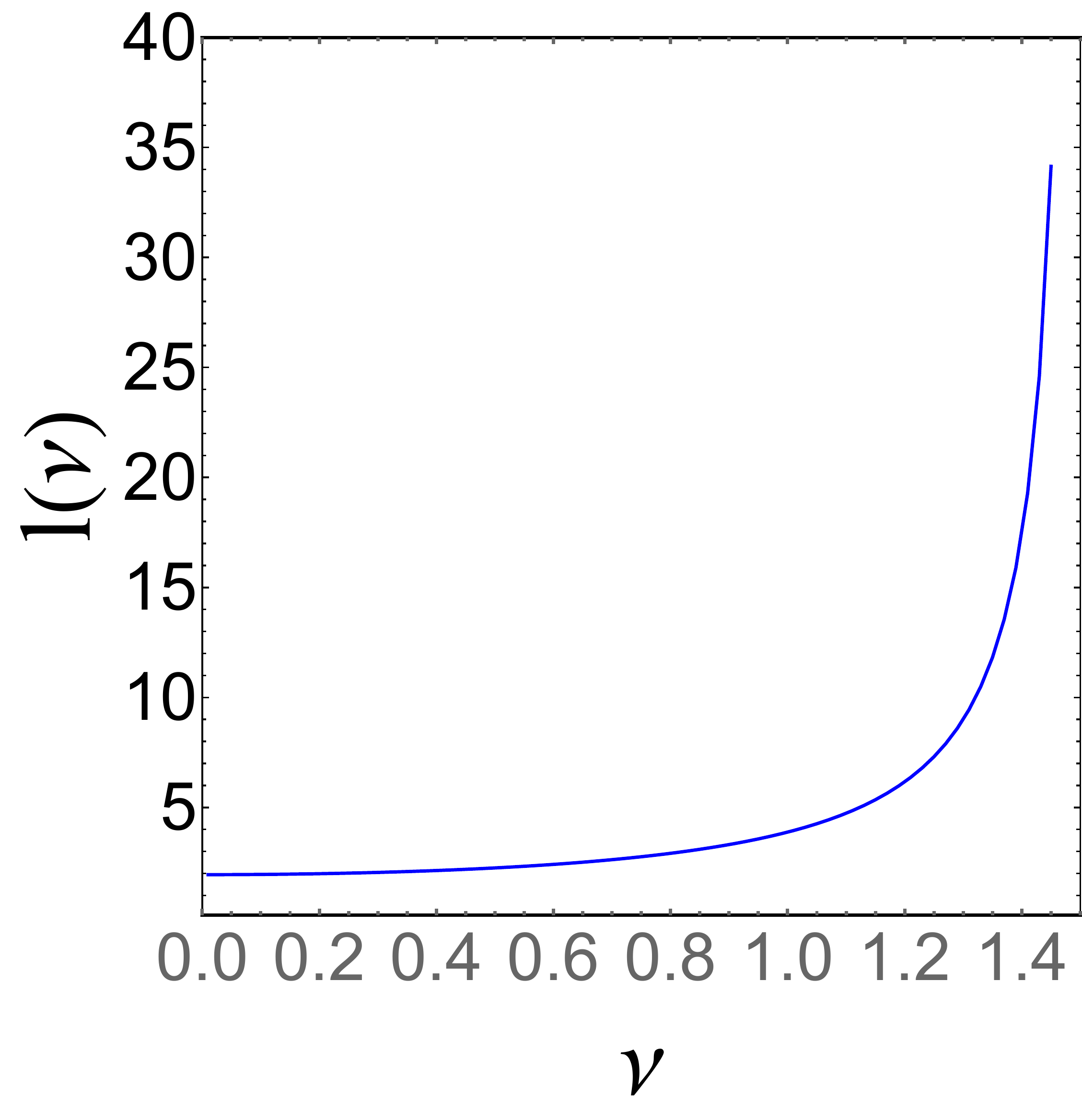}
	\caption{Numerical values of $g(\nu)$, $h(\nu)$, and $l(\nu)$. 
	 }
	\label{order1coeficient}
\end{figure}

Finally, we summarize the scalar power spectrum and bispectrum of this model (we refer the readers to the original paper~\cite{Chen:2009zp} for details). First, the power spectrum of the curvature perturbation $\zeta$, is given by
\begin{align}
\langle\zeta_{\mathbf k_1}\zeta_{\mathbf k_2}\rangle
=(2\pi)^3\delta^3(\mathbf k_1 + \mathbf k_2)\frac{2\pi^2}{k_1^3}P_{\zeta}~, ~~
P_{\zeta} = \frac{H^4}{(2\pi)^2r_0^2\dot{\theta}_0^2}\left[1+8\pa{\frac{\dot{\theta}_0}{H}}^2\mathcal C(\nu)\right]\,,
\end{align}
where $\nu=\sqrt{\frac{9}{4}-\frac{m^2}{H^2}}$ and the formulae provided in this subsection are for $m<\frac{3}{2}H$, i.e., for real $\nu$. Also $\mathcal C(\nu)$ is a mass-dependent numerical coefficient defined by\footnote{See Ref.~\cite{Chen:2012ge} for its analytic form in terms of special functions. See also Ref.~\cite{Pi:2012gf} for a related work.}
\begin{align}
\mathcal C(\nu) 
&= \frac{\pi}{4}\,\mbox{Re}\left[\int^\infty_0\frac{dx_1}{x_1^{1/2}}\int^\infty_{x_1}\frac{dx_2}{x_2^{1/2}}
\left( H_\nu^{(1)}(x_1)e^{ix_1}H_\nu^{(2)}(x_2)e^{-ix_2}
\right.\right.\nn\\&\left.\left.\qquad\qquad\qquad\qquad\qquad\qquad\qquad
-H_\nu^{(1)}(x_1)e^{-ix_1}H_\nu^{(2)}(x_2)e^{-ix_2}\right)\right],
\end{align}
where $H^{(1)}(z)$ and $H^{(2)}(z)$ are Hankel functions of the first  and second kinds, respectively. As we mentioned, $\dot\theta_0/H$ characterizes the mixing of the adiabatic and isocurvature modes\footnote{Note that the linear mixing term $\dot{\varphi}\sigma$ has less derivatives than the kinetic term, so that when $\sigma$ is massless, it is dominant at the superhorizon regime. It leads a divergence of $\mathcal{C}(\nu)$ at $\nu=3/2$ (see Fig.2 in Ref.~\cite{Chen:2012ge}). For the same reason, $g(\nu)$, $h(\nu)$, and $l(\nu)$ diverge at $\nu=3/2$ as shown in Fig.~\ref{order1coeficient}.}.   

\medskip
The nonlinearity parameter $f_{\rm NL}$, characterizing bispectra is defined such that
\begin{align}
\label{defNG}
	\langle \zeta_{\mathbf k_1}\zeta_{\mathbf k_2}\zeta_{\mathbf k_3} \rangle\to (2\pi)^7\delta^3(\mathbf k_1+\mathbf k_2+\mathbf k_3) P_{\zeta}^2 \bigg(\frac{9}{10}f_{\rm NL}\bigg) \frac{1}{k_1^2k_2^2k_3^2}
\end{align}
in the equilateral limit $k_1=k_2=k_3$. Each cubic coupling in Eq.~\eqref{action_varphi} sources non-Gaussianities of the magnitude~\cite{Chen:2009zp},
\begin{align} 
f_{\rm NL}^{\sigma^3} 
= -g(\nu) \lambda_3\pa{\frac{\dot{\theta}_0}{H}}^4\,,
\quad f_{\rm NL}^{\sigma^2\varphi} 
= -h(\nu)  \pa{\frac{\dot{\theta}_0}{H}}^4,\quad
f_{\rm NL}^{\sigma\varphi^2} = -l(\nu) \left(\frac{\dot \theta_0}{H}\right)^2\,.
\label{fNL}
\end{align}
Here $g(\nu)$, $h(\nu)$ and $l(\nu)$ are mass-dependent numerical coefficients defined by (see also Fig.~\ref{order1coeficient} for their numerical values)
\begin{align}
g(\nu) &= \frac{5\pi^3}{48}  \bigg(\frac{m}{H}\bigg)^2  {\rm Im} \int_0^\infty \frac{dx_1}{x_1^4} I (x_1) I  ( x_1 ) I  (  x_1 ) ~,
\\
h(\nu) &=   \frac{5\pi^2   }{12 } {\rm Im} \int_{0}^{\infty} dx_1 \frac{e^{-i x_1}}{x_1^2}    I ( x_1 ) I ( x_1 ) ~, 
\\
l(\nu) &= -\frac{5 \pi}{6} {\rm Im} \int_{0}^{\infty}  dx_1e^{-2i x_1} I (x_1)  + \frac{ 5\pi  }{12} {\rm Im} \int_{0}^{\infty} \frac{dx_1}{x_1^2}e^{-2i x_1} (1+i x_1)^2  I ( x_1 )  \,,
\end{align}
with $I(x_1)$ being defined by
\begin{align}
I(x_1)&= x_1^{3 / 2}\left[2 \operatorname{Im}\left[ {H}_{\nu}^{(1)}(x_1) \int_{0}^{\infty} \frac{{d} x_2}{x_2^{1/2}}  {H}_{\nu }^{(2)}\left(x_2\right) e^{-i x_2}\right]\right.\\
\nonumber
&\quad
\left.+ {iH}_{\nu}^{(1)}(x_1) \int_{0}^{x_1} \frac{{d} x_2}{x_2^{1/2}} {H}_{\nu }^{(2)}\left(x_2\right) e^{- i x_2}-i  {H}_{\nu }^{(2)}(x_1) \int_{0}^{x_1} \frac{{d} x_2}{x_2^{1/2}}  {H}_{\nu}^{(1)}\left(x_2\right) e^{- i x_2}\right]\,.
\end{align}
Note that $f_{\rm NL}^{\sigma\varphi^2}$ and $f_{\rm NL}^{\sigma^2\varphi}$ are small in the perturbative regime, $\dot{\theta_0}\lesssim H$, of the linear mixing. On the other hand, the cubic self-coupling $\sigma^3$, of the isocurvature mode may generate observable non-Gaussianities with $|f_{\rm NL}^{\sigma^3}| \ge\mathcal O(1)$, {\it if} $\lambda_3$ is large enough to compensate the suppression by $\dot{\theta}_0/H$.

\subsection{Non-Gaussianities vs renormalizability}
\label{subsec:examples}

We have reviewed that in quasi-single field inflation non-Gaussianities with $|f_{\rm NL}|>\mathcal{O}(1)$ may be sourced by the cubic self-coupling $\lambda_3$ if it is large enough. In the following we discuss if such a large non-Gaussianity is realized consistently.
To illustrate our motivation, we end this section by providing a simple observation that large non-Gaussianities cannot be realized if the cubic coupling is originated from a renormalizable Mexican hat potential.

\medskip
In the Cartesian coordinates of the field space, the $U(1)$ symmetric renormalizable potential is parameterized by two parameters, $\lambda$ and $r_{\rm min},$ as
\begin{align}
\mathcal{V}_{U(1)}=\frac{\lambda}{2} (\phi_1^2+\phi_2^2-r_{\rm min}^2)^2\,,
\end{align}
and so the potential of the radial coordinate, $r$, is
\begin{align}
V_r(r)=\frac{\lambda}{2}(r^2-r_{\rm min}^2)^2\,.
\end{align}
Then, the effective potential~\eqref{V_sigma} of the isocurvature mode $\sigma$ reads
\begin{align}
	V(\sigma)&=V_r(r_0+\sigma)-V_r(r_0)-\left[
\frac{\dot{\theta}_0^2}{2}(r_0+\sigma)^2-\frac{\dot{\theta}_0^2}{2}r_0^2\right]
\nonumber\\
&=\frac{m^2r_0^2}{2}\left[ \left(\frac{\sigma}{r_0}\right)^2+\left(\frac{\sigma}{r_0}\right)^3+\frac{1}{4}\left(\frac{\sigma}{r_0}\right)^4 \right]\,,
\label{U1potential}
\end{align} 
where we used the equation of motion~\eqref{eom_r} and defined the isocurvature mass $m^2=4\lambda r_0^2$. Note that $\lambda_3$ and $\lambda_4$ are no longer free parameters, but rather they are specified as $\lambda_3=\lambda_4=1$. In particular, the nonlinearity parameter $f_{\rm NL}^{\sigma^3}\sim (\dot{\theta}_0/H)^4$, is too small to observe at least in the perturbative regime of the linear mixing.

\medskip
This simple observation suggests that it is hard to obtain observable non-Gaussianities in two-field quasi-single field inflation without turning on nonrenormalizable interactions. In the following, we study perturbative unitarity of scattering amplitudes and clarify under which conditions observable non-Gaussianities are realized in quasi-single field inflation.

\section{Implications of perturbative unitarity}\label{Perturbative}

Based on the aforementioned motivation, we study high energy scattering of fluctuations around inflationary backgrounds and discuss implications of perturbative unitarity. For later convenience, we slightly generalize the model~\eqref{action_varphi} in the previous section as\footnote{
Note that derivatives of $\varphi$ always appear in the combination, $-2r_0\dot{\theta}_0 \dot{\varphi}
+(\partial_\mu\varphi)^2$, because of the Lorentz invariance of the original action for backgrounds. Also this form of action is determined by the symmetry of fluctuations alone, independent of details of the background dynamics. See, e.g., Ref.~\cite{Noumi:2012vr}.
}
\begin{align}
S&=\int dtd^3xa^3\bigg[
-\frac{1}{2}(\partial_\mu\varphi)^2
-\frac{1}{2}(\partial_\mu\sigma)^2-V(\sigma)
\nonumber
\\
&\qquad\qquad\qquad\quad
-\frac{\sigma}{r_0}\left[
-2r_0\dot{\theta}_0 \dot{\varphi}
+
(\partial_\mu\varphi)^2
\right]
-\frac{\alpha}{2}
\frac{\sigma^2}{r_0^2}
\left[
-2r_0\dot{\theta}_0 \dot{\varphi}
+
(\partial_\mu\varphi)^2
\right]
\bigg]
\\
&=\int dtd^3xa^3\bigg[
-\frac{1}{2}(\partial_\mu\varphi)^2
-\frac{1}{2}(\partial_\mu\sigma)^2-V(\sigma)
\nonumber
\\
\label{model_perturbations}
&\qquad\qquad\qquad\quad
+2\dot{\theta}_0\dot{\varphi}\sigma
+\alpha\frac{\dot{\theta}_0}{r_0}\dot{\varphi}\sigma^2
-\frac{\sigma}{r_0}(\partial_\mu\varphi)^2
-\frac{\alpha}{2}
\frac{\sigma^2}{r_0^2}
(\partial_\mu\varphi)^2
\bigg]
\,,
\end{align}
which reproduces Eq.~\eqref{action_varphi} for $\alpha=1$. Also we use the same parameterization~\eqref{V_sigma} of the isocurvature potential.
In the rest of the section, we begin by a brief review of the unitarity bound (Sec.~\ref{perturbativeunitarity}). Then, in Sec.~\ref{ImplicationQSI} we discuss implications of perturbative unitarity for the model~\eqref{model_perturbations}. There we demonstrate that $\alpha=\lambda_3=\lambda_4=1$ and so small non-Gaussianities, $|f_{\rm NL}|\lesssim 1$, are required if we decouple gravity and assume that the model is UV complete at the tree-level. This confirms our simple observation in the previous section. In Sec.~\ref{scale} we generalize the argument to include theories with a finite cutoff scale $\Lambda$. We provide a relation between the nonlinearity parameter $f_{\rm NL}$ and the scale $\Lambda$ of new physics beyond quasi-single field inflation which is required for UV completion.

\subsection{Unitarity bound}\label{perturbativeunitarity}

The starting point is the $S$-matrix unitarity:
\begin{align}
SS^\dag =\mathbb 1
\,.
\label{smatrixunitarity0}
\end{align}
Using the transition matrix $\mathcal T$ defined by
\begin{align}
S = \mathbb 1 + i \mathcal T\,,
\end{align}
we write Eq.~\eqref{smatrixunitarity0} as 
\begin{align}
-i(\mathcal T - \mathcal T^\dag) = \mathcal T\mathcal T^\dag~.\label{TTdag}
\end{align}
Next we rewrite it in term of the scattering amplitude:
\begin{align}
\bra{B}\mathcal T\ket{A} = (2\pi)^4\delta^4(p_A - p_B)M_{AB}\,,\label{Tmatrix}
\end{align}
where $|A\rangle$ and $|B\rangle$ are the initial and final states, respectively. Also $p_A$ is the total momentum of the initial state $|A\rangle$ and similarly for $p_B$.
In this language, Eq.~\eqref{TTdag} reads
\begin{align}
&-i(2\pi)^4\delta^4(p_A - p_B)(M_{AB} - M^*_{BA}) 
\nn\\
&=\sum_C \prod_{i=1}^{N_C} \int \frac{d^{3}{\mathbf p}_i}{(2\pi)^3}\frac{1}{2E_i({\mathbf p}_i)} (2\pi)^4\delta^4(p_A - p_B)(2\pi)^4\delta^4(p_A - p_C) M_{CB}M^*_{CA}\,,
\label{SunitarityT}
\end{align}
where $N_C$ is the number of external particles in the intermediate state $C$ and $E_i({\mathbf p}_i)$ is the on-shell energy of the $i$-th particle with the spatial momentum ${\mathbf p}_i$.
In particular, for identical initial and final states, $A=B$, we have
\begin{align}
2\,\text{Im}M_{AA}
&=\sum_C \prod_{i=1}^{N_C} \int \frac{d^{3}{\mathbf p}_i}{(2\pi)^3}\frac{1}{2E_i({\mathbf p}_i)}(2\pi)^4\delta^4(p_A - p_C) |M_{CA}|^2\,.
\end{align}
Since each summand in the right-hand side is non-negative, we have
\begin{align}
\label{master}
2|M_{AA}|
&\geq\prod_{i=1}^{N_C} \int \frac{d^{3}{\mathbf p}_i}{(2\pi)^3}\frac{1}{2E_i({\mathbf p}_i)} (2\pi)^4\delta^4(p_A - p_C) |M_{CA}|^2\,,
\end{align}
which implies a bound on high-energy scattering with a typical energy scale $E$ as
\begin{align}
\label{M_ii_bound}
|M_{AA}|\leq E^{4-2N_A}\,.
\end{align}
Substituting this bound back into Eq.~\eqref{master}, we have
\begin{align}
\prod_{i=1}^{N_C} \int \frac{d^{3}{\mathbf p}_i}{(2\pi)^3}\frac{1}{2E_i({\mathbf p}_i)} (2\pi)^4\delta^4(p_A - p_C) |M_{CA}|^2\leq E^{4-2N_A}\,,
\end{align}
which implies a bound on more general high-energy scattering as
\begin{align}
|M_{CA}|\leq E^{4-(N_A+N_C)}\,.
\end{align}
For example, bounds on four-, five-, and six-point amplitudes read\footnote{
Note that the bounds presented here are applicable in four-dimensional spacetime. Bounds in general spacetime dimensions $d$ are given by
\begin{align}
\label{general_d}
|M_{\text{4pt}}|\leq E^{4-d}\,,
\quad
|M_{\text{5pt}}|\leq E^{5-\frac{3}{2}d}\,,
\quad
|M_{\text{6pt}}|\leq E^{6-2d}\,.
\end{align}}
\begin{align}
\label{unitarity_bounds}
|M_{\text{4pt}}|\leq E^0\,,
\quad
|M_{\text{5pt}}|\leq E^{-1}\,,
\quad
|M_{\text{6pt}}|\leq E^{-2}\,.
\end{align}
If we assume weakly coupled UV completion, these bounds have to be satisfied by tree-level amplitudes. Note that we have suppressed $\mathcal{O}(1)$ coefficients in the inequalities~\eqref{M_ii_bound}-\eqref{unitarity_bounds}, but they can be fixed in the standard manner, e.g., using angular momentum eigenstates. We will take care of these $\mathcal{O}(1)$ coefficients when necessary. See Appendix~\ref{cutoffscale} for details.

\subsection{Constraints from perturbative unitarity}
\label{ImplicationQSI}

Let us use the bounds~\eqref{unitarity_bounds} to constrain the model parameters $\alpha,\lambda_3,\lambda_4$ in two-field quasi-single field inflation~\eqref{model_perturbations} and discuss their implication to non-Gaussianities. In this subsection we focus on the case where gravity is decoupled and derive conditions for the model to be UV complete at the tree-level. 

\begin{figure}[t]
\centering
\includegraphics[width=3cm]{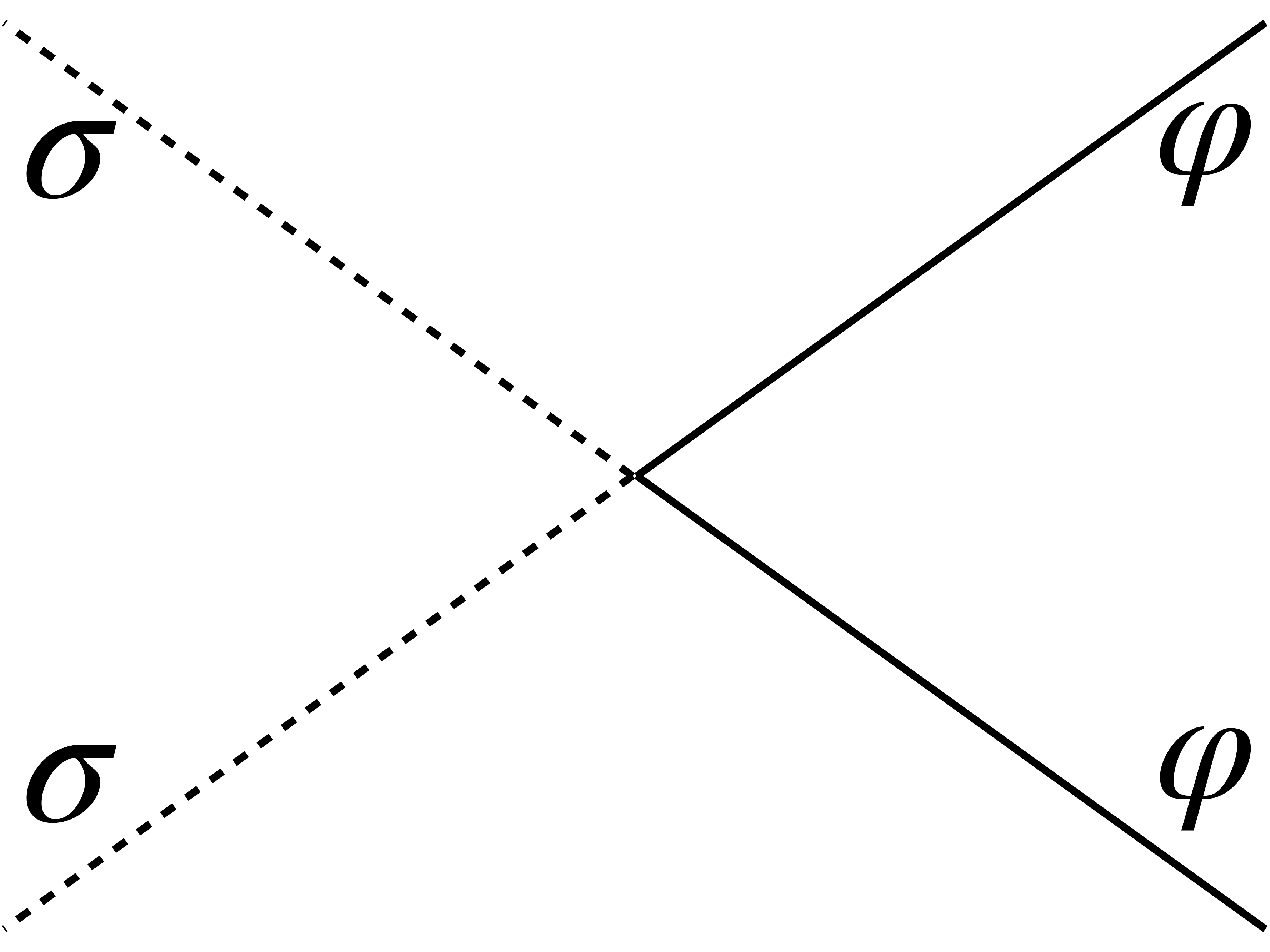}\qquad\qquad
\includegraphics[width=3cm]{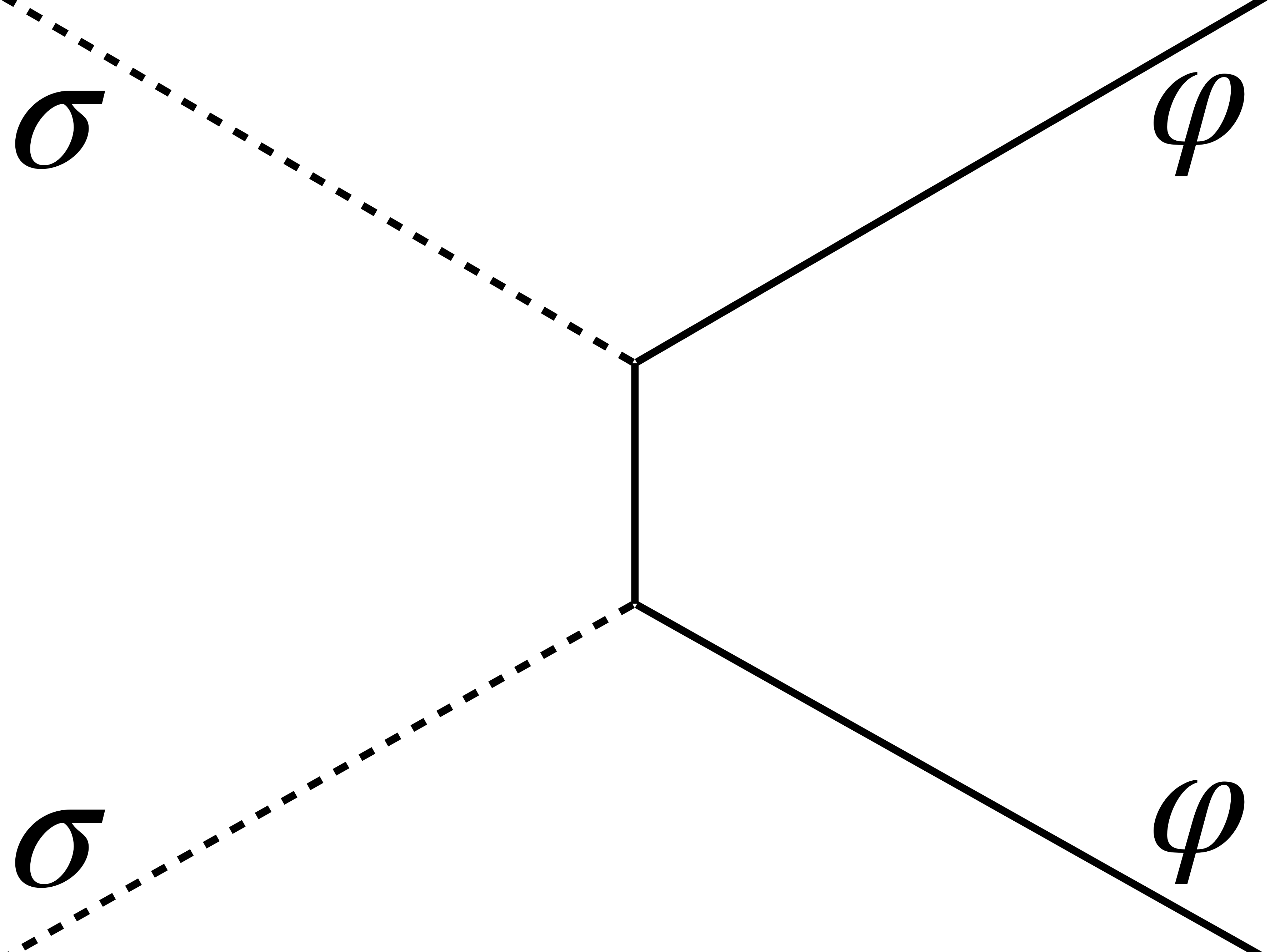}
\caption{Feynman diagrams which generate $E^2$ contributions in $\sigma\sigma\to\varphi\varphi$ scattering.}
\label{4ptfig}
\end{figure}

\paragraph{Four-point scattering.}

We begin by four-point scattering. Perturbative unitarity constraints on general scalar field theories were nicely discussed in Ref.~\cite{Nagai:2019tgi}. A conclusion there is that in spacetime four dimensions, the model is UV complete at the tree-level only when the field space is flat\footnote{
Note that in $d=2$ a nonzero internal space curvature is allowed since the unitarity bound is milder as shown in Eq.~\eqref{general_d}, which is consistent with the fact that 2d CFT accommodates a curved target space.}. In other words, the field space curvature gives a cutoff scale of the theory. In our model~\eqref{model_perturbations}, the field space curvature tensor at the origin $\sigma=\varphi=0$ is given in terms of the parameter $\alpha$ as
\begin{align}
\left.R_{\sigma\varphi\sigma\varphi}\right|_{\sigma=\varphi=0}=\frac{1-\alpha}{r_0^2}\,,
\end{align}
and similarly for other components. Then, a nontrivial constraint is available from $\sigma\sigma\varphi\varphi$ scattering\footnote{
When we need to specify which particles are in the initial/final states, we call scattering process, e.g., as $\sigma\sigma\to\varphi\varphi$ scattering, but otherwise we use the terminology such as $\sigma\sigma\varphi\varphi$ scattering. Correspondingly, we use, e.g., both of $M_{\sigma\sigma\varphi\varphi}$ and $M_{\sigma\sigma\to\varphi\varphi}$, to denote scattering amplitudes.}:
\begin{align}
M_{\sigma\sigma\varphi\varphi} &= 2(\alpha-1) \frac{p_3.p_4}{r_0^2} + \mathcal O(E^0) \,,
\label{4ptT2}
\end{align}
where we kept terms which grow up faster than $E^0$ at high energy $E\gg m,|\dot{\theta}_0|$. See Fig.~\ref{4ptfig} for the corresponding Feynman diagrams. Also  $p_i$ is the four-momentum (in the all-incoming notation) of the $i$-th external particle $\phi_i$ in the amplitude $M_{\phi_1\ldots\phi_n}$, and $p_i.p_j$ is the inner product of $p_i$ and $p_j$. Then, the unitarity bound \eqref{unitarity_bounds} implies
\begin{align}
\alpha &= 1~. \label{4ptunitaritylambda}
\end{align}

\begin{figure}[t]
\centering
\includegraphics[width=3cm]{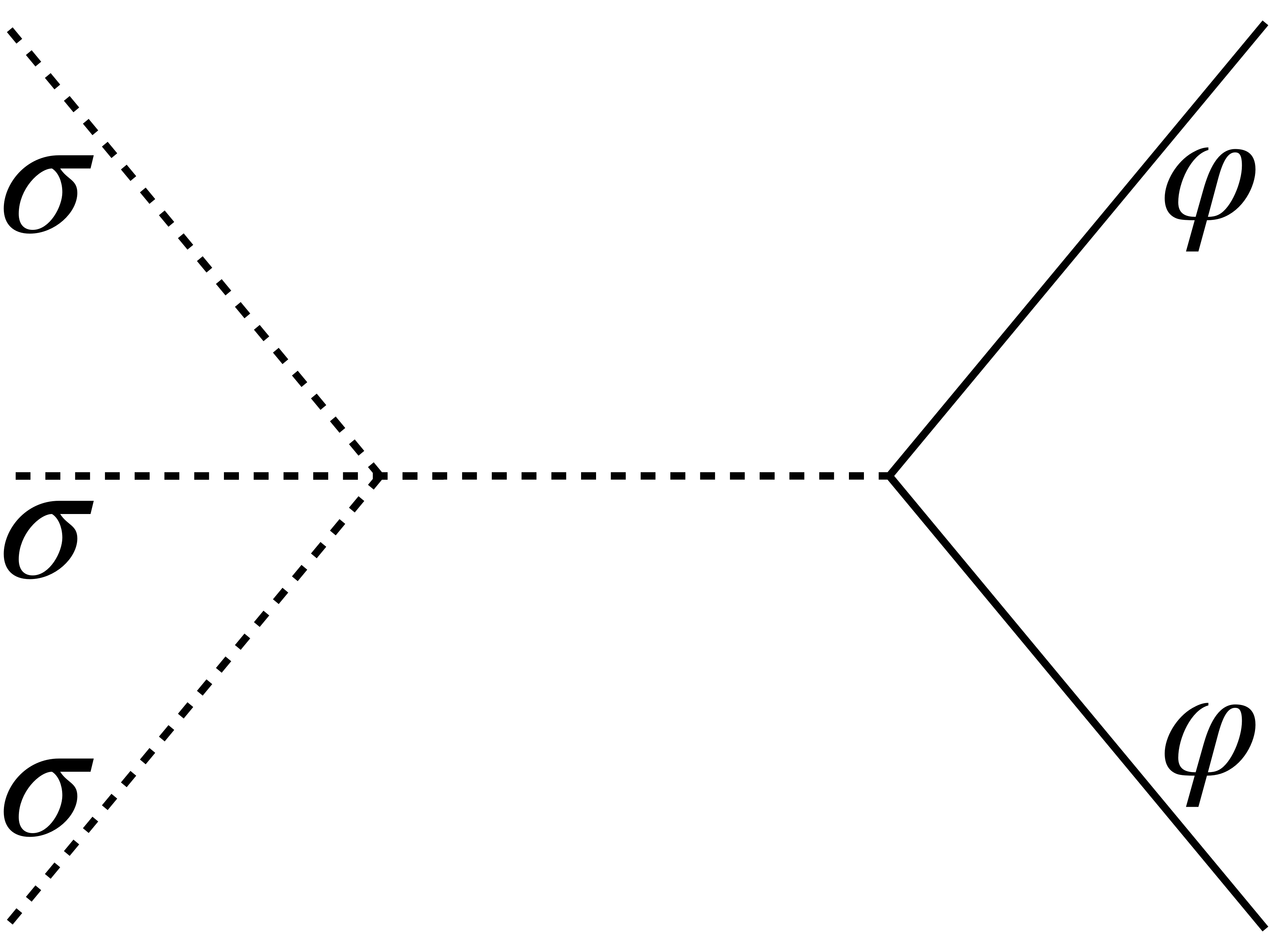}:(a)\qquad
\includegraphics[width=3cm]{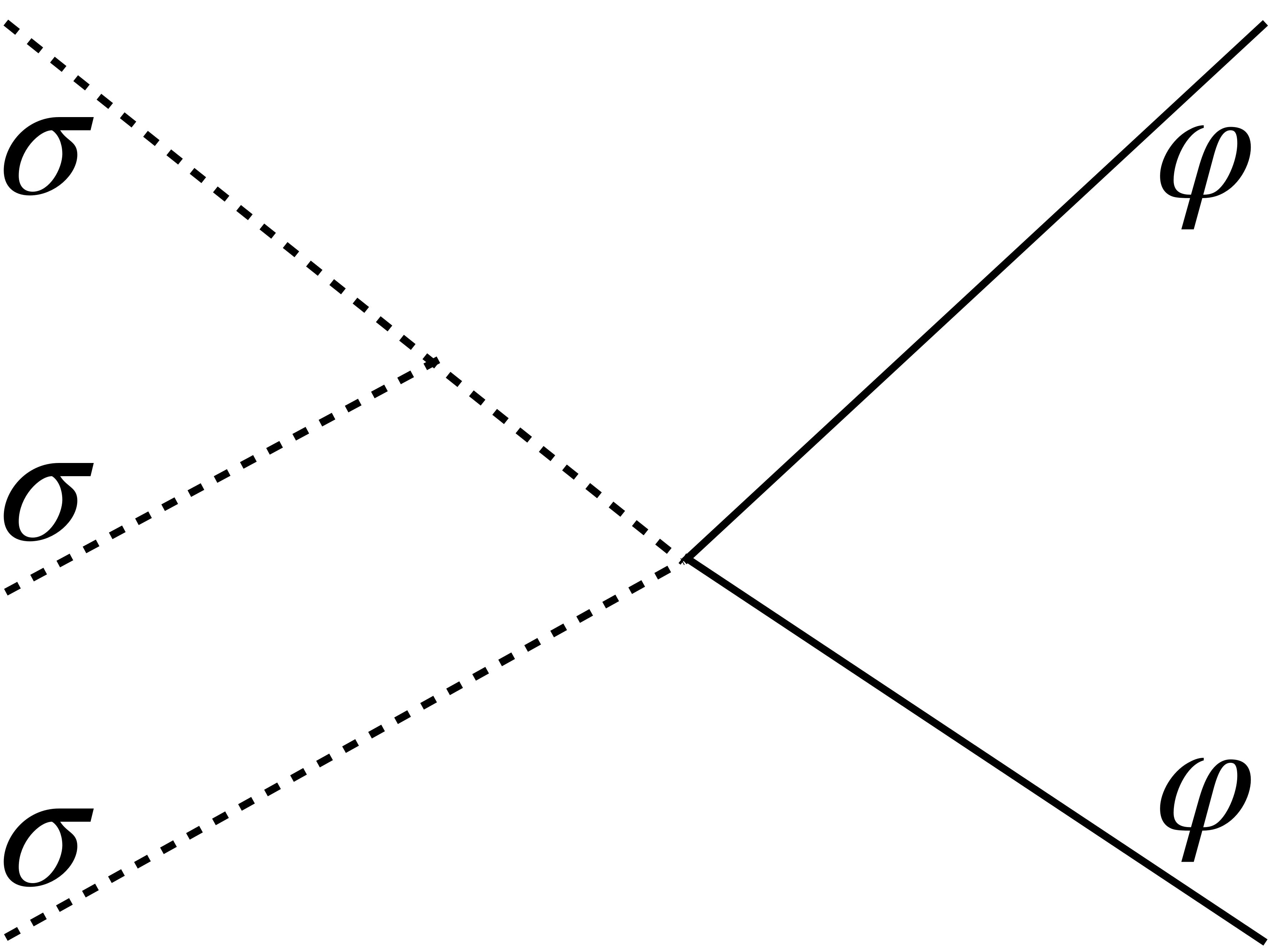}:(b)\qquad
\includegraphics[width=3cm]{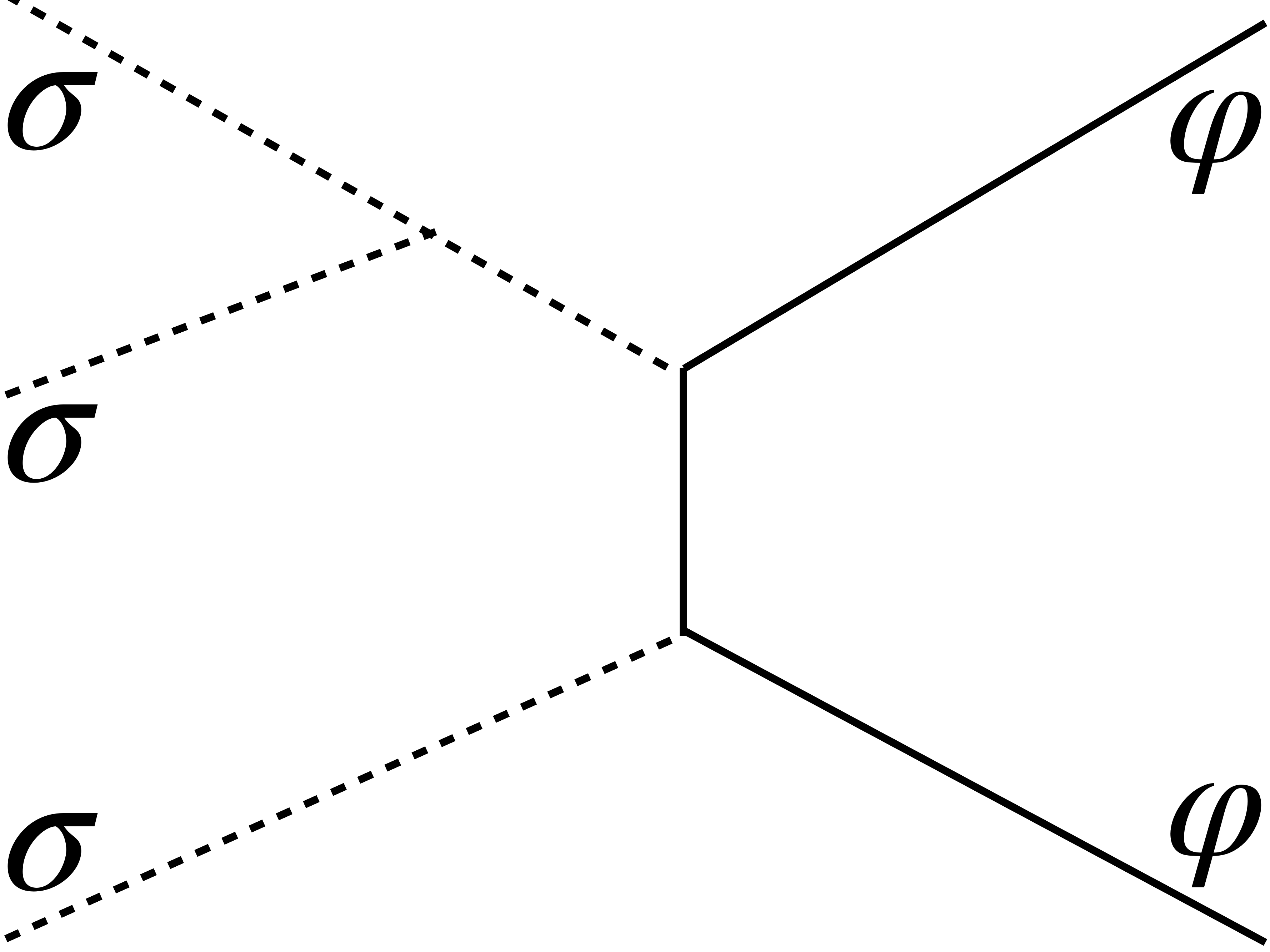}:(c)\\\vspace{0.5cm}
\includegraphics[width=3cm]{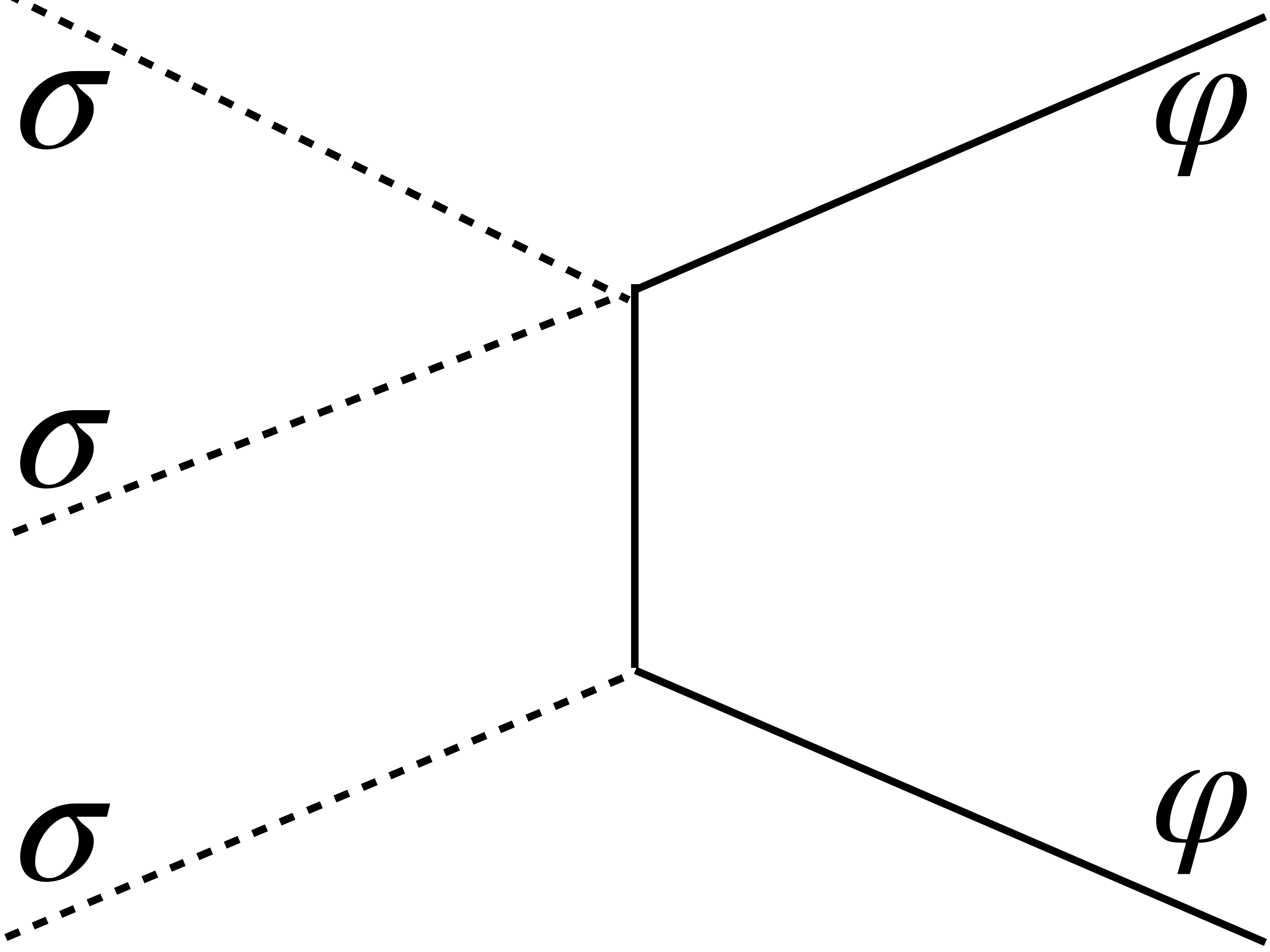}:(d)\qquad
\includegraphics[width=3cm]{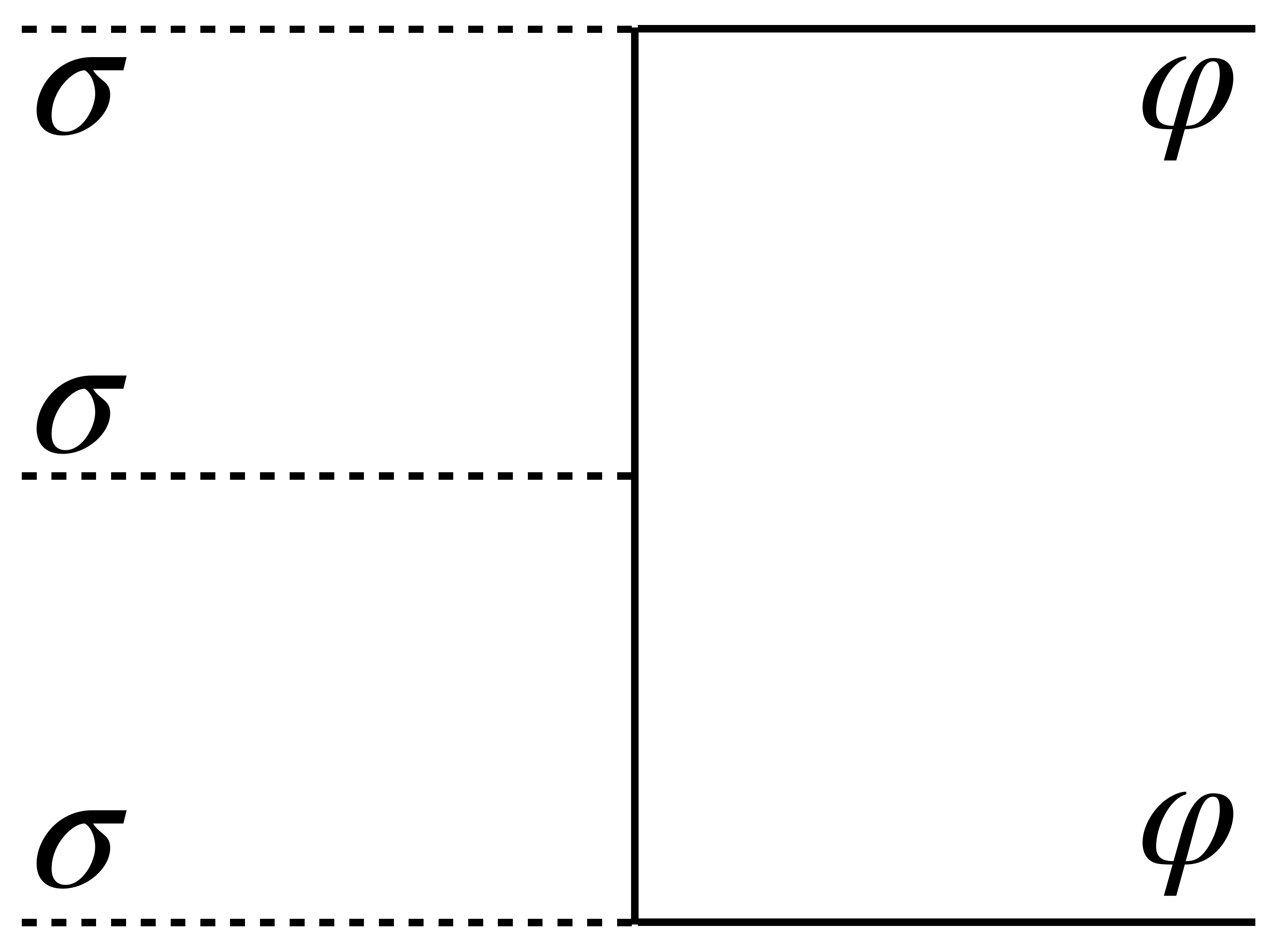}:(e)
\caption{The leading order Feynman diagrams of $\sigma\sigma\sigma\varphi\varphi$ scattering.} \label{fivepoint3sigma2phi}
\end{figure}

\begin{figure}[t]
\centering
\includegraphics[width=3cm]{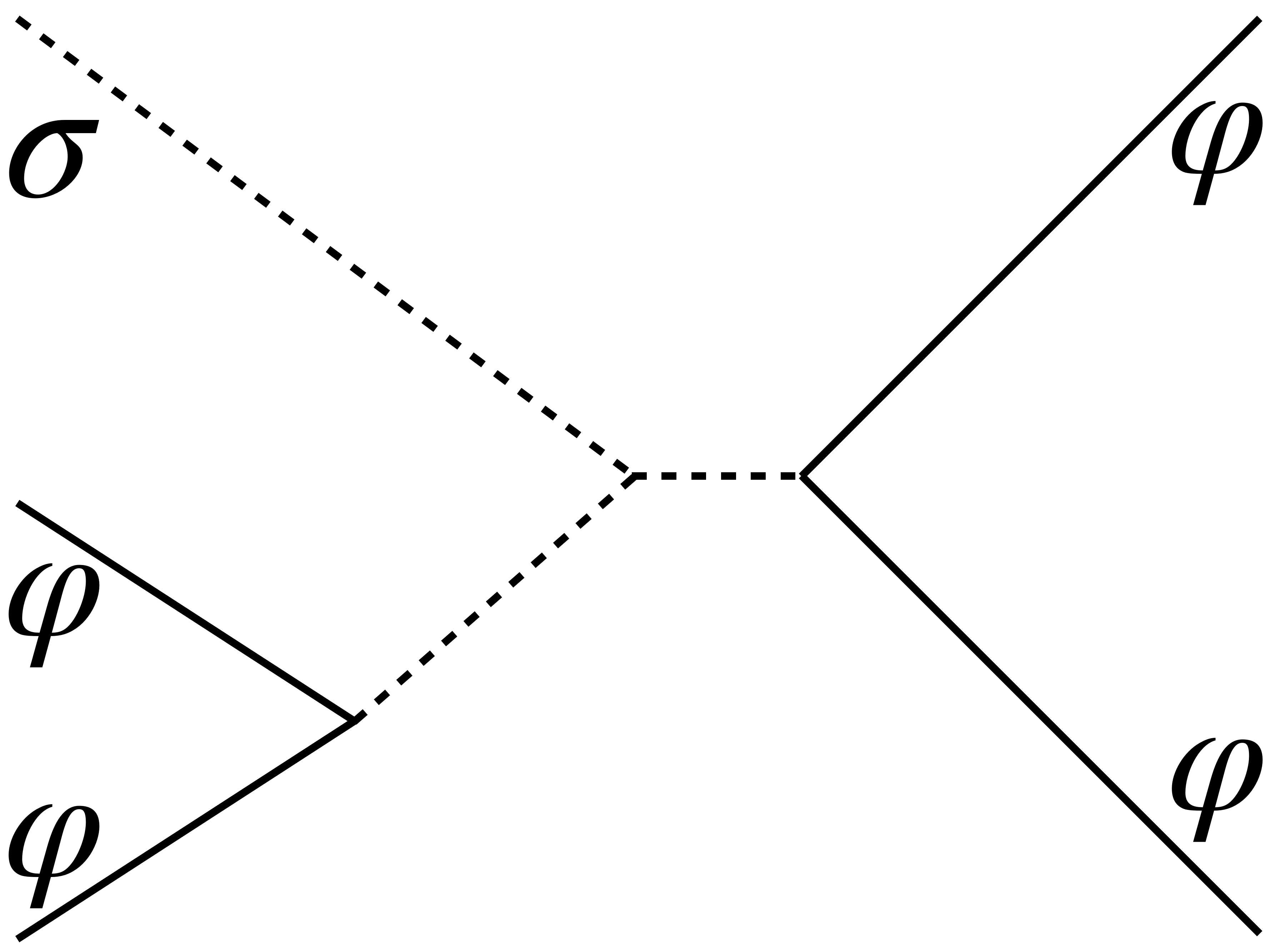}\qquad\qquad
\includegraphics[width=3cm]{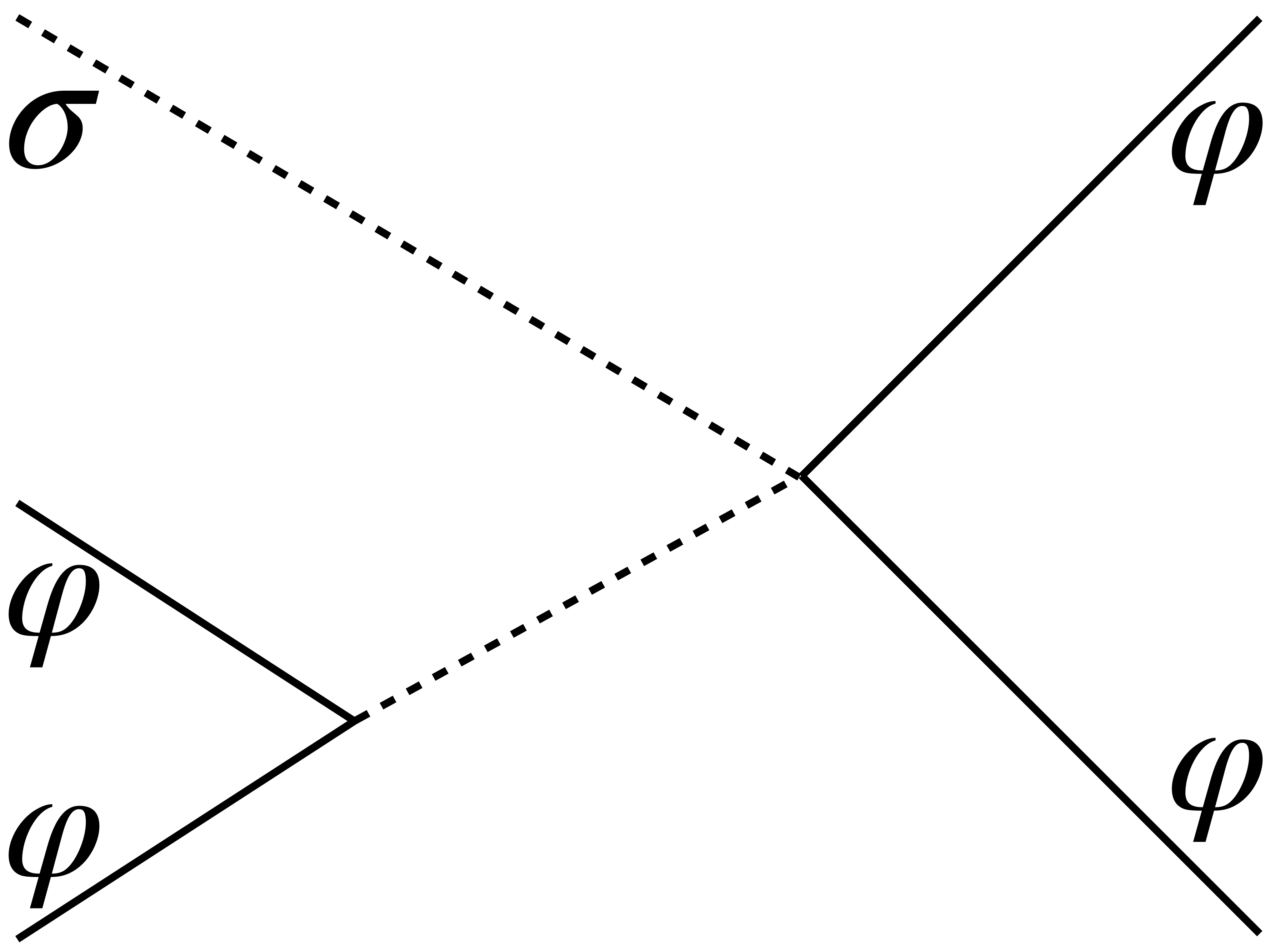}\qquad\qquad
\includegraphics[width=3cm]{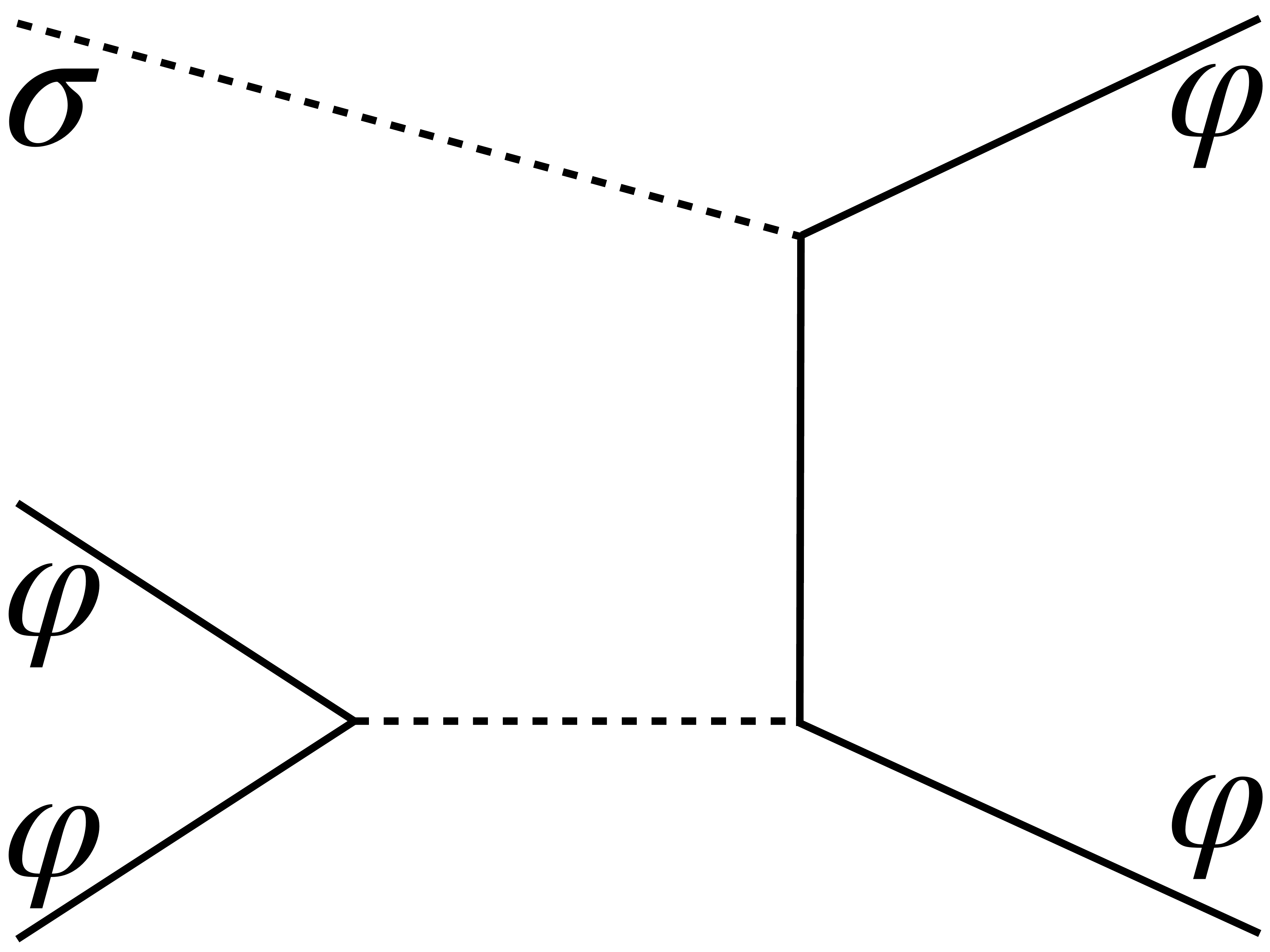}
\caption{The leading order Feynman diagrams of $\sigma\varphi\varphi\varphi\varphi$ scattering.}
\label{5ptspppp}
\end{figure}

\paragraph{Five-point scattering.}

Next we consider five-point scattering. In our model~\eqref{model_perturbations} there are two five-point scattering processes relevant for our purpose: $\sigma\sigma\sigma\varphi\varphi$ and $\sigma\varphi\varphi\varphi\varphi$.
The unitarity bound~\eqref{unitarity_bounds} requires that they are bounded by $\mathcal{O}(E^{-1})$. First, we consider the amplitude, $M_{\sigma\sigma\sigma\varphi\varphi}$. 
Its high energy behavior reads (see Fig.~\ref{fivepoint3sigma2phi} for the relevant diagrams)
\begin{align}
M_{\sigma\sigma\sigma\varphi\varphi} &= -8(\alpha-1)\left(\frac{p_4.p_5}{r_0^3}+\mathcal{O}(E^0)\right)
+ 3\pa{ 3\lambda_3 -\lambda_4-2}\frac{m^2}{r_0^3}  +\mathcal O(E^{-2})
\label{sigma3varphi2}
\,,
\end{align}
where the $\mathcal{O}(E^2)$ contribution generated by the diagrams (d) and (e) cancel each other out for $\alpha=1$ in particular (see also Appendix~\ref{app:redefinition} for field redefinition useful in the amplitude computation).
Now the unitarity bound \eqref{unitarity_bounds} implies
\begin{align}
\alpha=1\,,\quad 3\lambda_3-\lambda_4= 2 \,.\label{5ptunitaritylambda}
\end{align}
Similarly, the high energy behavior of $M_{\sigma\varphi\varphi\varphi\varphi}$ reads
\begin{align}
M_{\sigma\varphi\varphi\varphi\varphi} = -(\alpha-1)\frac{m^2}{r_0^3}\left(\frac{p_2.p_3}{p_4.p_5}+\text{5 perms.}\right)-\pa{\alpha+9\lambda_3-10}\frac{m^2}{r_0^3} + \mathcal O(E^{-2})\,.
\label{sig_phi4}
\end{align}
See Fig.~\ref{5ptspppp} for relevant diagrams.
Then, the unitarity bound \eqref{unitarity_bounds} implies
\begin{align}
\alpha=\lambda_3= 1\,. \label{unitaritybound3}
\end{align}
To summarize, perturbative unitarity of five-point scattering implies Eqs.~\eqref{5ptunitaritylambda} and \eqref{unitaritybound3}, so that we conclude that
\begin{align}
\label{UVC_lambda}
\alpha=\lambda_3=\lambda_4=1
\end{align}
when we decouple gravity and assume that the model is UV complete at the tree-level. Note that this precisely reproduces the renormalizable potential~\eqref{U1potential}\footnote{
Note that one may easily show that any $\mathcal{O}(\sigma^5)$ term in the potential, $V_\sigma$, violates the unitarity bound, so that the allowed potential is a quartic one with the conditions~\eqref{UVC_lambda}.}. As we discussed in the previous section, non-Gaussianities are too small to observe for this parameter set.
In other words, quasi-single field inflation cannot generate $|f_{\rm NL}|\gtrsim1$ without requiring a new physics. In the next subsection, we provide a relation between the nonlinearity parameter $f_{\rm NL}$ and the scale $\Lambda$ of such new physics beyond quasi-single field inflation.

\paragraph{Six-point scattering.}

\begin{figure}[t]
\centering
\includegraphics[width=3cm]{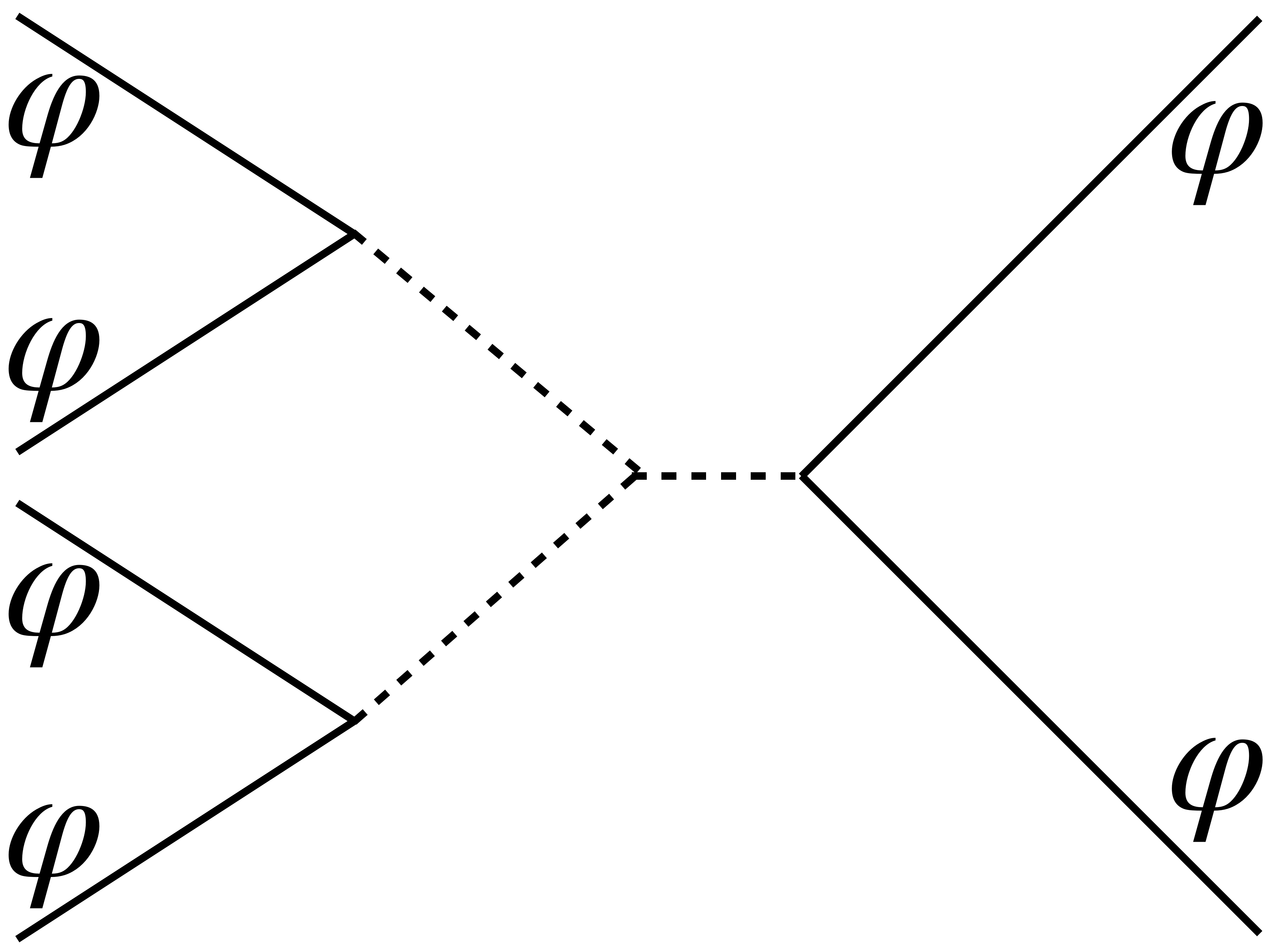}\qquad\qquad
\includegraphics[width=3cm]{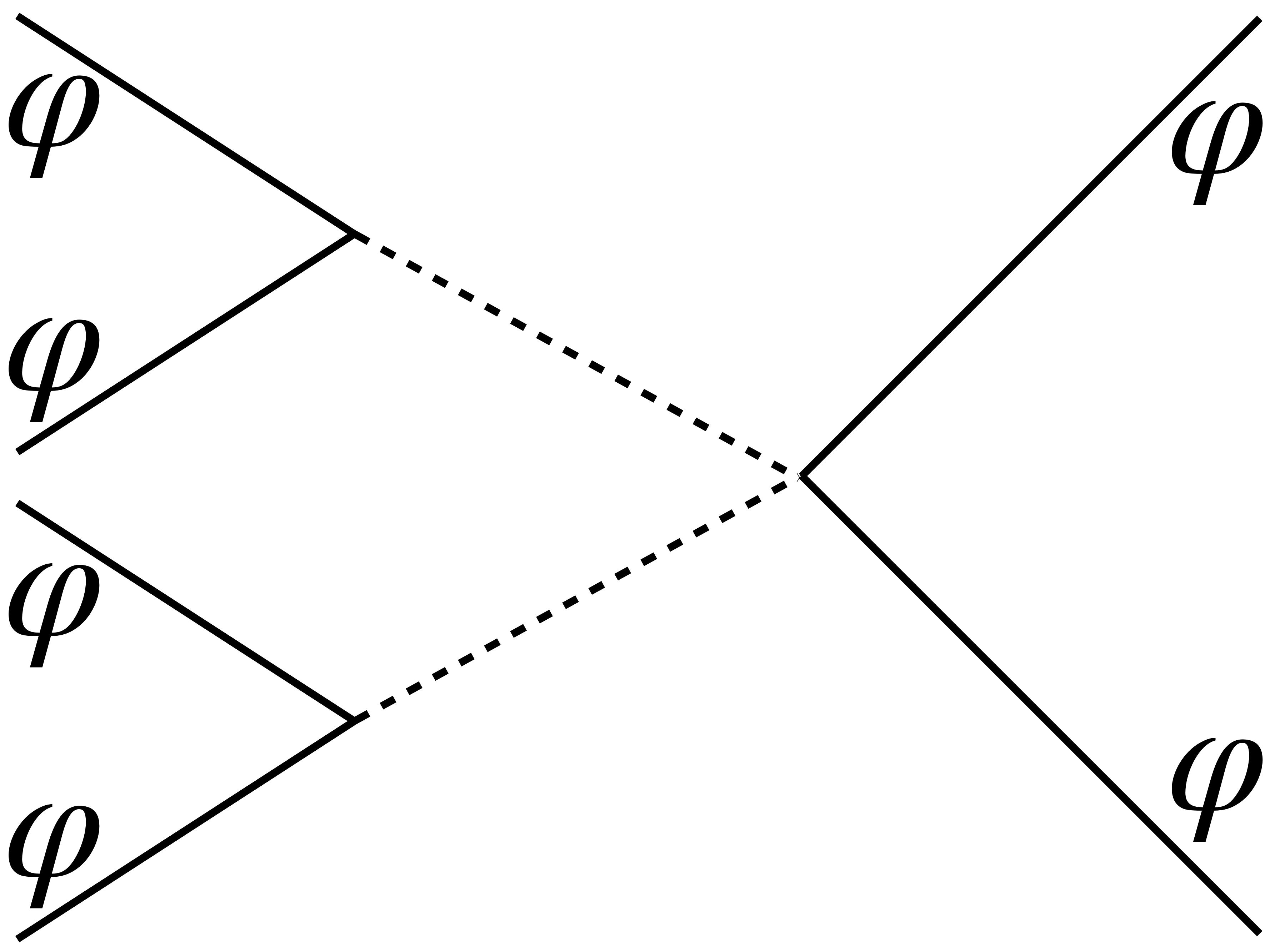}\qquad\qquad
\includegraphics[width=3cm]{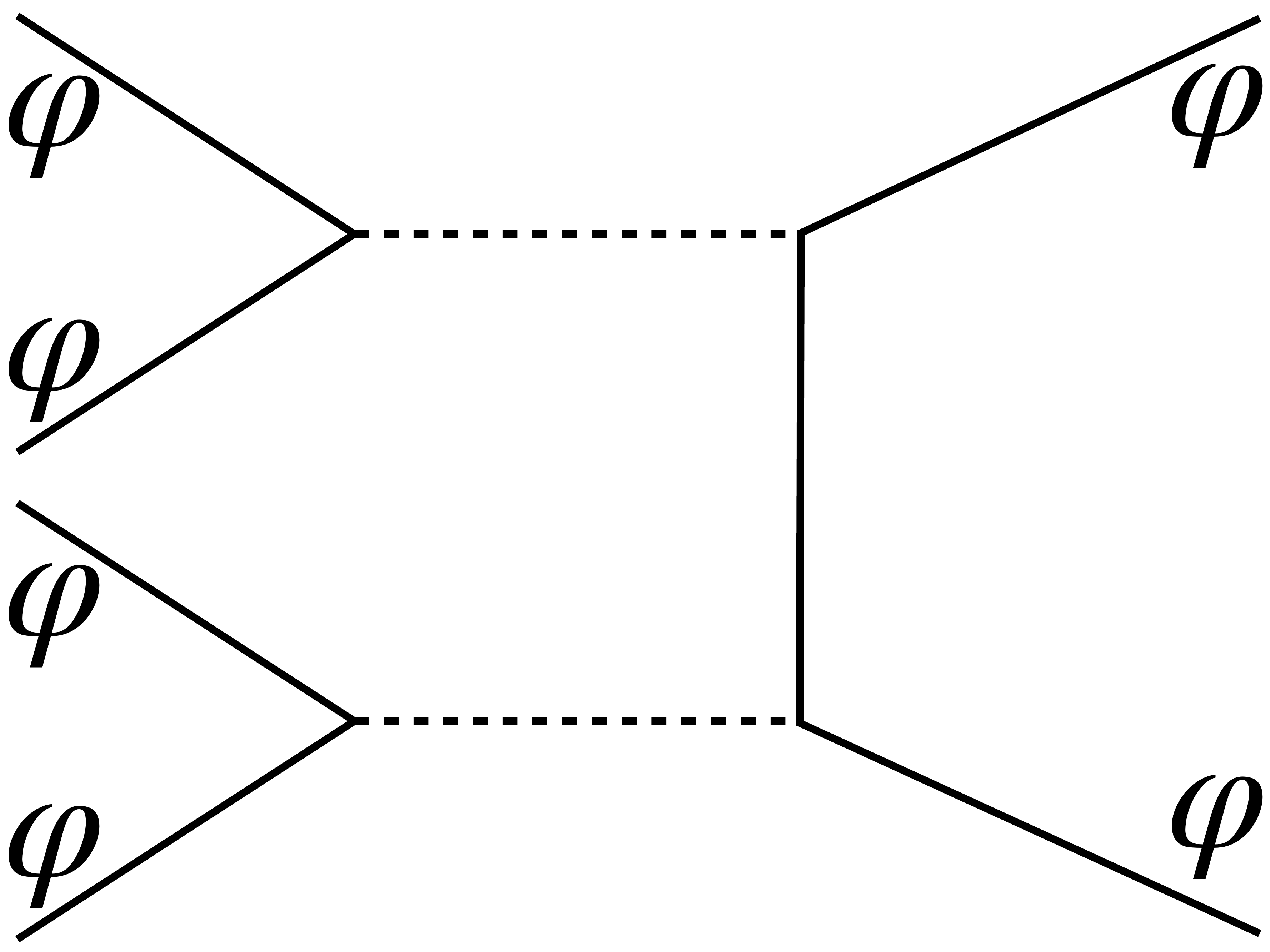}
\caption{The leading order Feynman diagrams of six-point scattering of $\varphi$.}
\label{6ptfig}
\end{figure}

Before proceeding to the next subsection, let us provide six-point scattering amplitudes of $\varphi$ for later convenience (see Fig.~\ref{6ptfig} for relevant diagrams):
\begin{align}
M_{\varphi\varphi\varphi\varphi\varphi\varphi} = -15\cdot (\alpha+3\lambda_3 - 4)\frac{m^2}{r_0^4}+\mathcal{O}(E^{-2})\,.\label{phi6}
\end{align}
We find that the $\mathcal{O}(E^0)$ term vanishes and so the unitarity bound is satisfied under the conditions~\eqref{UVC_lambda}.

\subsection{Scale of new physics}\label{scale}
In the previous subsection, we have shown that observable non-Gaussianities $|f_{\rm NL}|\gtrsim 1$ can be generated only when the perturbative unitarity is broken at some scale $\Lambda$. Therefore, if we assume a weakly coupled UV completion, the observable non-Gaussianities imply that the model has to be UV completed by a new additional particle with the mass $m\lesssim \Lambda$\footnote{
Another possibility is that the model becomes strongly coupled at the scale $\Lambda$ and it is UV completed in a non-perturbative manner, which typically implies a phase transition at the scale $\Lambda$ just like QCD.}.
In this subsection, we consider deviations from the UV complete parameter set $\alpha=\lambda_3=\lambda_4=1$ and discuss a relation between non-Gaussianities $f_{\rm NL}$ and the scale $\Lambda$ of new physics.

\medskip
In our model~\eqref{model_perturbations}, there are three types of cubic interactions, $\sigma^3$, $\sigma^2\dot{\varphi}$, and $\sigma(\partial_\mu\varphi)^2$. When the mixing between $\varphi$ and $\sigma$ is small at the Hubble scale ($|\dot{\theta}_0|\lesssim H$), each cubic coupling contributes to the nonlinearity parameter $f_{\rm NL}$ as
\begin{align}
\label{f_NL_sec3}
f_{\rm NL}^{\sigma^3} 
= -\lambda_3g(\nu)\pa{\frac{\dot{\theta}_0}{H}}^4\,,
\quad
f_{\rm NL}^{\sigma^2 \varphi} 
=  -\alpha h(\nu) \pa{\frac{\dot{\theta}_0}{H}}^4\,,
\quad
f_{\rm NL}^{\sigma \varphi^2} 
= -l(\nu) \pa{\frac{\dot{\theta}_0}{H}}^2\,,
\end{align}
where note that an overall factor $\alpha$ is multiplied  in the second expression compared to Eq.~\eqref{fNL} since we are allowing a nonzero field space curvature ($\alpha\neq1$). Our task is now to write down the cutoff scale $\Lambda$ derived from the perturbative unitarity in terms of the nonlinearity parameter $f_{\rm NL}$.

\paragraph{Deformations of isocurvature potential.}
\label{def_pot}

First, we consider the case when the cubic self-coupling $\lambda_3$ of the isocurvature mode $\sigma$ is the dominant source of non-Gaussianities. For simplicity, we assume $\lambda_3\neq1$, but $\alpha=1$.
As we discussed in the previous subsection, a deviation of $\lambda_3$ from unity breaks the perturbative unitarity of the five-point and six-point scattering amplitudes at a high-energy scale, which gives the cutoff scale of the theory. In appendix~\ref{cutoffscale}, we use unitarity of the $S$-wave amplitudes of $M_{\sigma\varphi\varphi\varphi\varphi}$ and $M_{\varphi\varphi\varphi\varphi\varphi\varphi}$ to derive the following cutoff scales taking care of numerical coefficients:
\begin{align}
\Lambda_{{\rm 5pt}}&=\frac{64\sqrt{2}\pi^2}{9}\cdot\frac{ r_0^3}{m^2|\lambda_3-1|}
\simeq 100\cdot\frac{ r_0^3}{m^2|\lambda_3-1|}\,,
\label{cutoff_5pt}
\\
\Lambda_{\rm 6pt}&=\frac{16\sqrt{2}\pi^{3/2} }{\sqrt{15}}\cdot\frac{r_0^2}{m\,|\lambda_3-1|^{1/2}}\simeq 33\cdot\frac{r_0^2}{m\,|\lambda_3-1|^{1/2}}
\,.
\label{cutoff_lambda3_pre}
\end{align}
Having in mind observable non-Gaussianities $|f_{\rm NL}|\gtrsim1$, let us assume that $\lambda_3$ is large enough to compensate the suppression by the factor $|\dot{\theta}_0/H|\lesssim1$. Then, by using Eq.~\eqref{f_NL_sec3}, the cutoff scales~\eqref{cutoff_5pt}-\eqref{cutoff_lambda3_pre} are translated as
\begin{align}
\Lambda_{{\rm 5pt}}&\simeq
4\times 10^{13}H\cdot \pa{\frac{2\times 10^{-9}}{P_\zeta}}^{3/2}\cdot\frac{|\dot\theta/H|}{0.1}\cdot
\left(\frac{\frac{H^2}{m^2}g(\nu)}{100}\right)\cdot
\left(\frac{1}{|f_{\rm NL}|}\right)\,,
\\
\Lambda_{\rm 6pt}&\simeq
4\times 10^{9}H\cdot \frac{2\times 10^{-9}}{P_\zeta}\cdot
\left(\frac{\frac{H^2}{m^2}g(\nu)}{100}\right)^{1/2}\cdot
\left(\frac{1}{|f_{\rm NL}|}\right)^{\frac{1}{2}}\,,
\label{cutoff_lambda3}
\end{align}
where note that $\frac{H^2}{m^2}g(\nu)\simeq 100$ for $m\simeq H$. We find that in the regime of our interests, unitarity of six-point amplitudes provide a stronger constraint.

\begin{figure}[t]
	\centering
	\includegraphics[width=9cm]{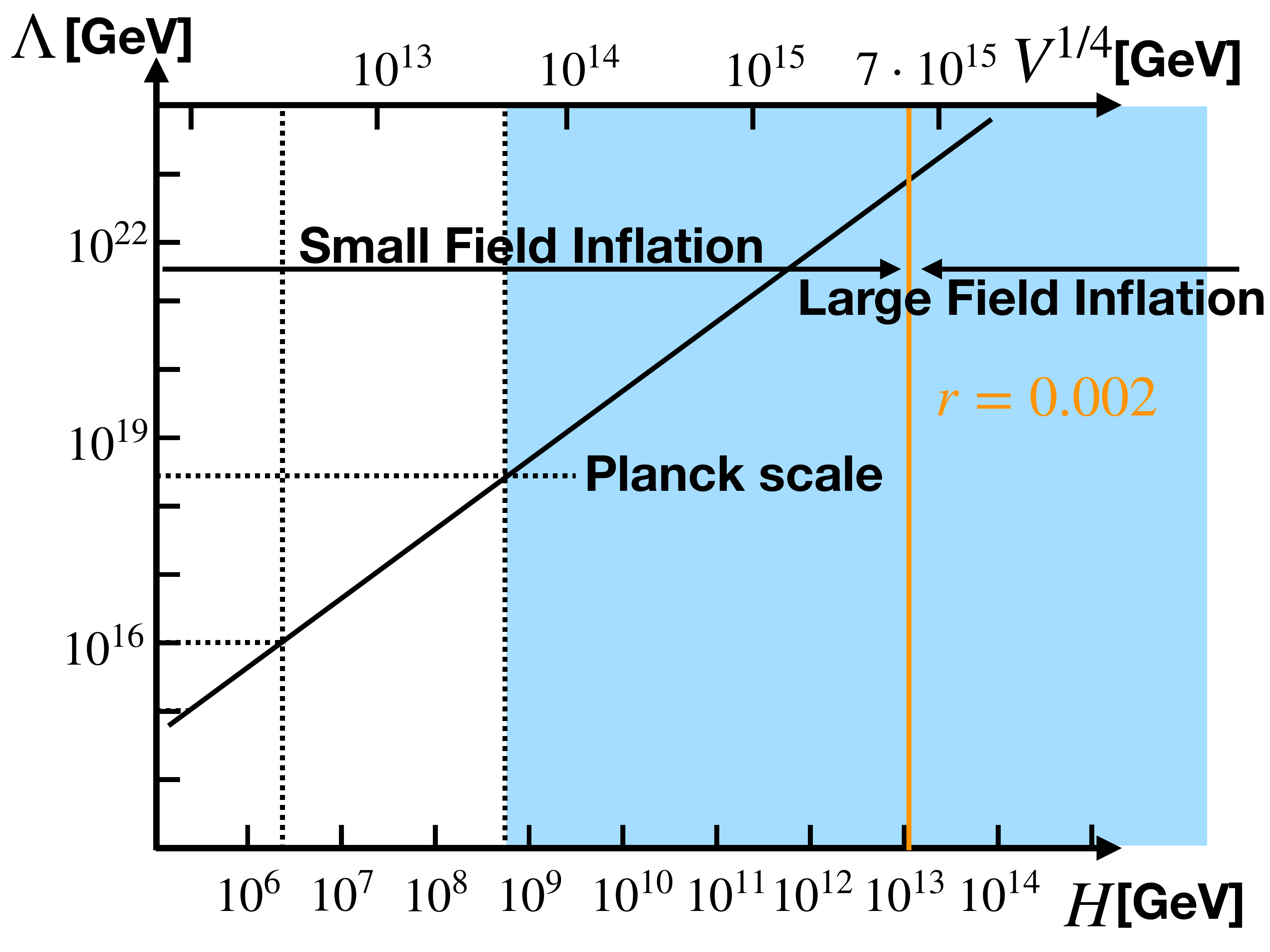}
	\caption{Cutoff scale $\Lambda_{\rm 6pt}$ vs Hubble scale $H$: As a benchmark point, we plot $\Lambda_{\rm 6pt}=4\times 10^9H$, which corresponds to the choice $\frac{H^2}{m^2}g(\nu)=100$ and $|f_{\rm NL}|=1$. The blue region corresponds to $\Lambda_{\rm 6pt}\ge M_{\rm pl}$.
	}
	\label{cutoff}
\end{figure}

\medskip
Now let us take a closer look at $\Lambda_{\rm 6pt}$. See Fig.~\ref{cutoff} for a plot of $\Lambda_{\rm 6pt}=4\times 10^9H$, taking $\frac{H^2}{m^2}g(\nu)=100$ and $|f_{\rm NL}|=1$ as a benchmark point. First, we find that the cutoff scale is near the Planck scale $M_{\rm Pl} \simeq 2.4\times 10^{18}$ GeV when $H\simeq 6\times 10^{8}$ GeV, or equivalently for the vacuum energy $V\simeq(5\times 10^{13}\,{\rm GeV})^4$. This implies that Planck suppressed operators can easily generate large non-Gaussanities when $H\gtrsim 6\times 10^{8}\,$GeV ($V^{1/4}\gtrsim 5\times 10^{13}$ GeV). In particular, for $H\gtrsim 6\times 10^{9}$ GeV ($V^{1/4}\gtrsim 2\times 10^{14}$ GeV), Planck suppressed operators generate too large non-Gaussianities $|f_{\rm NL}|\gtrsim100$ (assuming $\frac{H^2}{m^2}g(\nu)=100$) and so it is hard to realize successful quasi-single field inflation without introducing a mechanism to suppress quantum gravity corrections. Qualitatively, this is analogous to the Lyth bound~\cite{Lyth:1996im}, beyond which a mechanism to suppress quantum gravity effects is needed to accommodate a super-Planckian inflaton excursion. Notice however that our bound is stronger than the Lyth bound quantitatively. On the other hand, if the inflation scale is sufficiently low, $H\lesssim 6\times 10^{8}$ GeV ($V^{1/4}\lesssim 5\times 10^{13}\,$GeV), non-Gaussianities with $|f_{\rm NL}|\gtrsim 1$ require a new physics below the Planck scale. For example, we have a cutoff scale $\Lambda_{\rm 6pt}\simeq 1\times 10^{16}$ GeV for $H\simeq 3\times 10^{6}$ GeV ($V^{1/4}\simeq 4\times 10^{12}$ GeV) and $|f_{\rm NL}|=1$.

\paragraph{Curved field space.}
\label{curv_int}
\begin{figure}[tb]
\centering
\includegraphics[width=9cm]{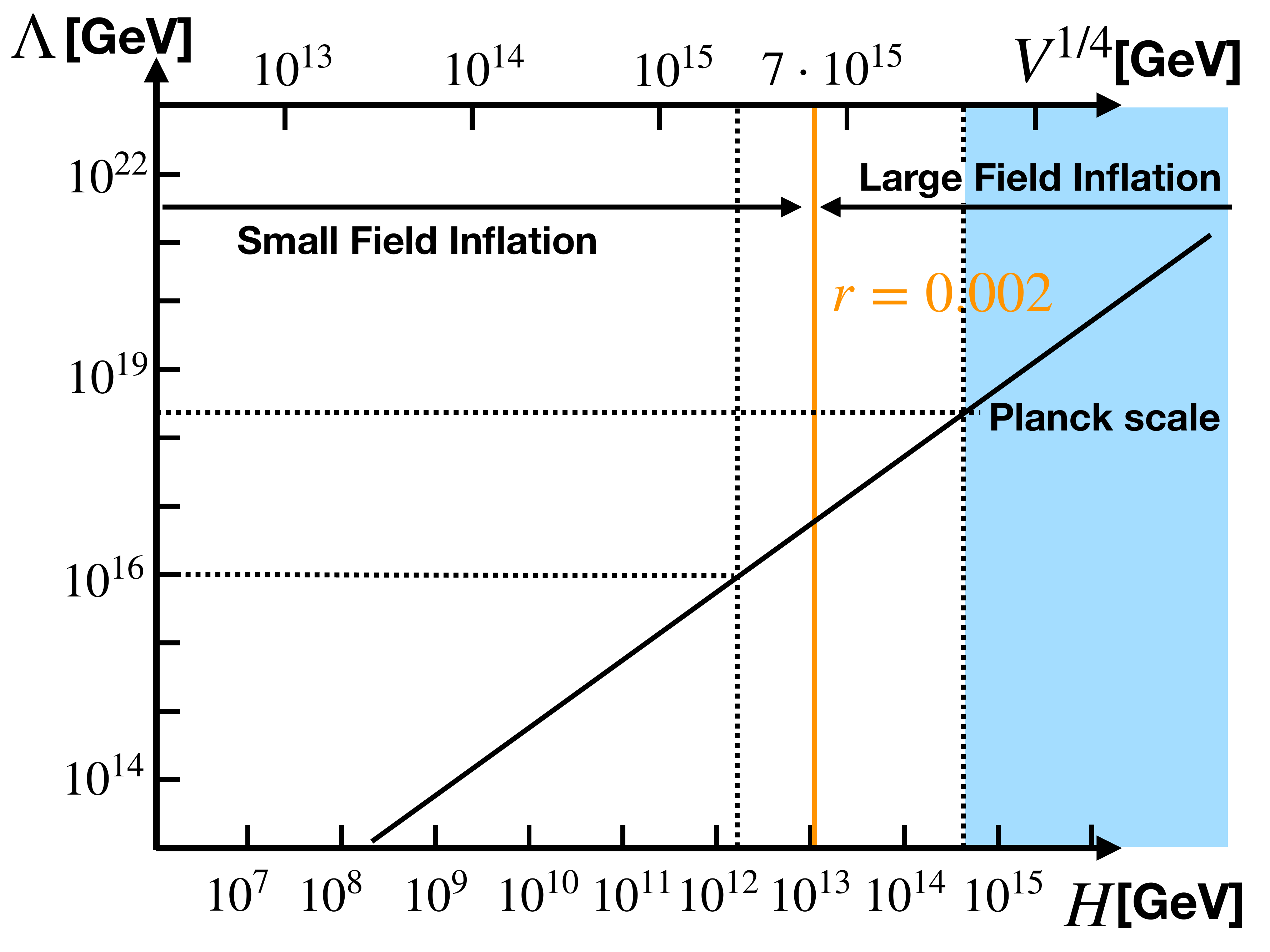}
\caption{
Cutoff scale $\Lambda_{\rm 4pt}$ vs Hubble scale $H$: As a benchmark point, we plot $\Lambda_{\rm 4pt}=6\times 10^3H$, which corresponds to the choice $|\dot{\theta}_0/H|=0.1$, $h(\nu)=10$, and $|f_{\rm NL}|=1$. The blue region corresponds to $\Lambda_{\rm 4pt}\ge M_{\rm pl}$.
}
\label{cutoff2}
\end{figure} 
Let us move on to the case with a curved field space $\alpha\neq1$. For simplicity, we assume $\lambda_3=1$, but the argument is qualitatively the same even when the cubic self-coupling of $\sigma$ generates non-Gaussianities comparable to the field space curvature effects $\lambda_3\simeq \alpha$. In the regime of our interests, unitarity of four-point scattering provides the strongest condition among the constraints discussed in the previous subsection:
\begin{align}
\label{cutoff_curvature_pre}
\Lambda_{\rm 4pt} =2\sqrt{2\pi}\cdot\frac{r_0}{|\alpha - 1|^{1/2}}  \simeq 5\cdot\frac{r_0}{|\alpha - 1|^{1/2}}
\,,
\end{align}
where the numerical coefficient was fixed by the strong coupling scale of the $S$-wave scattering. See Appendix~\ref{cutoffscale} for details. Having in mind observable non-Gaussianities $|f_{\rm NL}|\gtrsim1$, let us assume that $\alpha$ is large enough to compensate the suppression by the factor $|\dot{\theta}_0/H|\lesssim1$. Then, by using Eq.~\eqref{f_NL_sec3}, the cutoff scale~\eqref{cutoff_curvature_pre} is translated as
\begin{align}
\Lambda_{\rm 4pt} \simeq 6 \times 10^3 H\cdot
\left(\frac{2\times 10^{-9}}{P_\zeta}\right)^{1/2}\cdot
\frac{|\dot{\theta}_0/H|}{0.1}\cdot
\left(\frac{h(\nu)}{10}\right)^{1/2}\cdot
\left(\frac{1}{|f_{\rm NL}|}\right)^{\frac{1}{2}}\,,
\label{f_NL_curvature}
\end{align}
where note that $h(\nu)\simeq 10$ for $m\simeq H$. See Fig.~\ref{cutoff2} for a plot of $\Lambda_{\rm 4pt}=6\times 10^3H$, taking $|\dot{\theta}_0/H|=0.1$, $h(\nu)=10$ and $|f_{\rm NL}|=1$ as a benchmark point. We find that the cutoff scale is near the Planck scale $M_{\rm Pl}\simeq 2.4\times 10^{18}$ GeV for $H\simeq 4\times 10^{14}$ GeV ($V^{1/4}\simeq 4\times 10^{16}\,$GeV). Notice that this scale corresponds to the tensor-to-scalar ratio $r\simeq 1.6$ and so it is already ruled out by observations~\cite{Tristram:2020wbi}. Therefore, if the field space curvature effect $\alpha$ is a dominant source of non-Gaussianities, $|f_{\rm NL}|\gtrsim1$ requires a new physics below the Planck scale. For example, we have $\Lambda_{\rm 4pt}\simeq 1\times 10^{16}$ GeV for $H\simeq 2\times10^{12}$ GeV and $|f_{\rm NL}|=1$.

\bigskip
To summarize, we have shown that observable non-Gaussianities with $|f_{\rm NL}|\gtrsim 1$ require a new physics beyond quasi-single field inflation at some UV scale $\Lambda$. The relation between $f_{\rm NL}$ and the cutoff scale $\Lambda$ is summarized in Eq.~\eqref{cutoff_lambda3} and Eq.~\eqref{f_NL_curvature}. In particular, if the Hubble scale is $H\gtrsim 10^{9}$ GeV, a Planck suppressed $\sigma^3$ coupling can easily generate too large non-Gaussianities and so some mechanism to suppress quantum gravity effects is needed to realize a successful quasi-single field inflation.

\section{Heavy mass regime}\label{heavymass}

In the previous section, we focused on the case when the isocurvature mass is comparable to the Hubble scale $m\sim H$, which is the mass regime of quasi-single field inflation.  In this section, on the other hand, we consider the heavy mass regime $m\gg H$ and study the effective theory after integrating out the isocurvature mode $\sigma$. Our primary interest in this section is in classifying the parameter space of the $P(X,\phi)$ model,
\begin{align}
S=\int d^4x\sqrt{-g}\left[\frac{M^2_{\rm pl}}{2}R+P(X,\phi)\right]\,,
\end{align}
based on properties of the UV theory behind. Here $\phi$ is the inflaton, $X=-\frac{1}{2}(\partial_\mu \phi)^2$, and $P(X,\phi)$ is a function of $X$ and $\phi$. Note that the model is equivalent to the EFT of inflation with a unitarity gauge Lagrangian of the form $\mathcal{L}=\mathcal{L}(\delta g^{00},t)$~\cite{Cheung:2007st}. In general, the Lagrangian  contains higher derivative interactions, which source non-Gaussianities, and so it is interpreted as an effective Lagrangian of the inflaton after integrating out UV degrees of freedom. Then, the function $P(X,\phi)$ contains information of the UV theory, such as masses, spins, and the number of heavy fields. Below, we identify the EFT parameter space scanned by the UV theory with the inflaton and a single heavy isocurvature mode.

\subsection{UV complete parameter set ($\alpha=\lambda_3=\lambda_4=1$)}

First, we study the heavy mass regime of the UV complete parameter set $\alpha=\lambda_3=\lambda_4=1$. As we discussed, this parameter set is equivalent to the quasi-single field inflation model
\begin{align}
S&=\int d^4 x\sqrt{-g}\left[\frac{M_{\rm Pl}^2}{2}R-\frac12r^2 \del_\mu\theta\del^\mu\theta - \frac12  \del_\mu r\del^\mu r - V_r(r)- V_{\rm soft}(\theta)\right]\,,
\end{align}
with the isocurvature potential
\begin{align}
V_r(r)=\frac{\lambda}{2}(r^2-r_{\rm min}^2)^2\,.
\end{align}
To derive the inflaton effective action, we neglect the kinetic term of $r$ (since the isocurvature mode is frozen) and complete the square with respective to $r^2$ as
\begin{align}
S&\simeq\int d^4 x\sqrt{-g}\Bigg[\frac{M_{\rm Pl}^2}{2}R
-\frac12r_{\rm min}^2 \del_\mu\theta\del^\mu\theta
+\frac{1}{8\lambda}(\partial_\mu\theta\partial^\mu\theta)^2
- V_{\rm soft}(\theta)
\nonumber
\\
&\qquad\qquad\qquad\quad
-\frac{\lambda}{2}\left(r^2+\frac{1}{2\lambda}(\partial_\mu\theta\partial^\mu\theta)-r_{\rm min}^2\right)^2
\Bigg]\,.
\end{align}
Integrating out $r$ and defining $\phi=r_{\rm min}\theta$ gives the following inflaton effective action:
\begin{align}
\label{EFT_X^2}
S_{\rm eff}&=
\int d^4 x\sqrt{-g}\left[\frac{M_{\rm Pl}^2}{2}R
+X+\frac{1}{2\lambda r_{\rm min}^4}X^2
- V_{\rm soft}(\phi/r_{\rm min})\right]\,,
\end{align}
where $X=-\frac{1}{2}(\partial_\mu \phi)^2$ as before. We find that for the UV complete parameter set $\alpha=\lambda_3=\lambda_4=1$, there appear no $\mathcal{O}(X^3)$ terms in the inflaton effective action after integrating out the isocurvature mode. In other words, the $P(X,\phi)$ model with non-vanishing $\mathcal{O}(X^3)$ terms cannot be UV completed by a single heavy scalar in the weakly coupled regime. Note that the only approximation we have used here is that the isocurvature mode is heavy $m\gg H$. In particular, we have not made any assumption on the size of the mixing between the inflaton fluctuation and the isocurvature mode.

\medskip
It is also useful to rephrase our observation in terms of the EFT of inflation~\cite{Cheung:2007st}. Under the slow-roll approximation, the effective action of the Nambu-Goldstone boson $\pi$ is given in the decoupling limit as
\begin{align}
\label{EFToI}
S=\int dtd^3xa^3
\left[
-M_{\rm Pl}^2\dot{H}\left(\dot{\pi}^2-\frac{(\partial_i\pi)^2}{a^2}\right)
+\sum_{n=2}^\infty \frac{M_n^4}{n!}\left(-2\dot{\pi}-\dot{\pi}^2+\frac{(\partial_i\pi)^2}{a^2}\right)^n
\right]\,.
\end{align}
Here we neglected terms which contain second and higher order derivatives of $\pi$, which is the same assumption as the $P(X,\phi)$ model. In this language, the effective action~\eqref{EFT_X^2} is
\begin{align}
M_{\rm Pl}^2\dot{H}=-\frac{r_{\rm min}^2\dot{\theta}_0^2}{2}
\left(1+\frac{\dot{\theta}_0^2}{2\lambda r_{\rm min}^2}\right)
\,,
\quad
M_2^4=\frac{\dot{\theta}_0^4}{4\lambda }
\,,
\quad
M_n^4=0\quad(n\geq3)\,,
\end{align}
where the Nambu-Goldstone boson $\pi$ and the fluctuation $\delta\theta$ of the angular variable $\theta$ are related with each other as $\delta\theta=\dot{\theta}_0\pi$ under the slow-roll approximation. In terms of $r_0$ and $m$ defined in Eq.~\eqref{eom_r} and Eq.~\eqref{V_sigma} (see also Sec.~\ref{subsec:examples}), we may also write
\begin{align}
M_{\rm Pl}^2\dot{H}=-\frac{r_0^2\dot{\theta}_0^2}{2}
\,,
\quad
M_2^4=\frac{r_0^2\dot{\theta}_0^4}{m^2}
\,,
\quad
M_n^4=0\quad(n\geq3)\,.
\end{align}

\subsection{General values of $\alpha$, $\lambda_3$, $\lambda_4$}

We have seen that the $P(X)$ model with non-vanishing $\mathcal{O}(X^3)$ terms cannot be UV completed by a single heavy scalar in the weakly coupled regime. A natural question to ask is at which scale we need new physics beyond the two-field UV model (the UV model with an inflaton and a heavy scalar) when the $P(X)$ model contains $\mathcal{O}(X^3)$ terms.

\medskip
To answer this question, we start with the two-field model~\eqref{model_perturbations} and assume that the isocurvature mode $\sigma$ is heavy $m\gg H$. Having non-Gaussianity phenomenology in mind, let us rewrite the action~\eqref{model_perturbations} in terms of the Nambu-Goldstone boson $\pi=\varphi/(r_0\dot{\theta}_0)$ as
\begin{align}
S&=\int dtd^3xa^3\bigg[
-\frac{r_0^2\dot{\theta}_0^2}{2}(\partial_\mu\pi)^2
-\frac{1}{2}(\partial_\mu\sigma)^2-V(\sigma)
\nonumber
\\
&\qquad\qquad\qquad\,\,
-r_0\dot{\theta}_0^2\sigma\left[
-2 \dot{\pi}-\dot{\pi}^2+\frac{(\partial_i\pi)^2}{a^2}
\right]
-\frac{\alpha \dot{\theta}_0^2}{2}
\sigma^2
\left[
-2 \dot{\pi}-\dot{\pi}^2+\frac{(\partial_i\pi)^2}{a^2}
\right]
\bigg]\,,
\end{align}
with the isocurvature potential
\begin{align}
V(\sigma)=m^2r_0^2\left[
\frac{1}{2}\left(\frac{\sigma}{r_0}\right)^2+\frac{\lambda_3}{2}\left(\frac{\sigma}{r_0}\right)^3
+\frac{\lambda_4}{8}\left(\frac{\sigma}{r_0}\right)^4
+\mathcal{O}(\sigma^5)
\right]\,.
\end{align}
Assuming that cubic and higher order interactions are in the perturbative regime (its validity is discussed at the end of the section), we may easily integrate out the isocurvature $\sigma$ to obtain the effective action~\eqref{EFToI} with the EFT parameters (see, e,g, Ref.~\cite{Tolley:2009fg,Gong:2013sma}),
\begin{align}
\label{EFT_parameters}
M_{\rm Pl}^2\dot{H}=-\frac{r_0^2\dot{\theta}_0^2}{2}
\,,
\quad
M_2^4=\frac{r_0^2\dot{\theta}_0^4}{m^2}
\,,
\quad
M_3^4=3\left(\lambda_3-\alpha\right)\frac{r_0^2\dot{\theta}_0^6}{m^4}\,,
\end{align}
and similarly for $M_n^4$ ($n\geq4$). As expected, $M_3^4$ vanishes for the UV complete parameter set $\alpha=\lambda_3=\lambda_4=1$.

\begin{figure}[t]
	\centering
	\includegraphics[width=9cm]{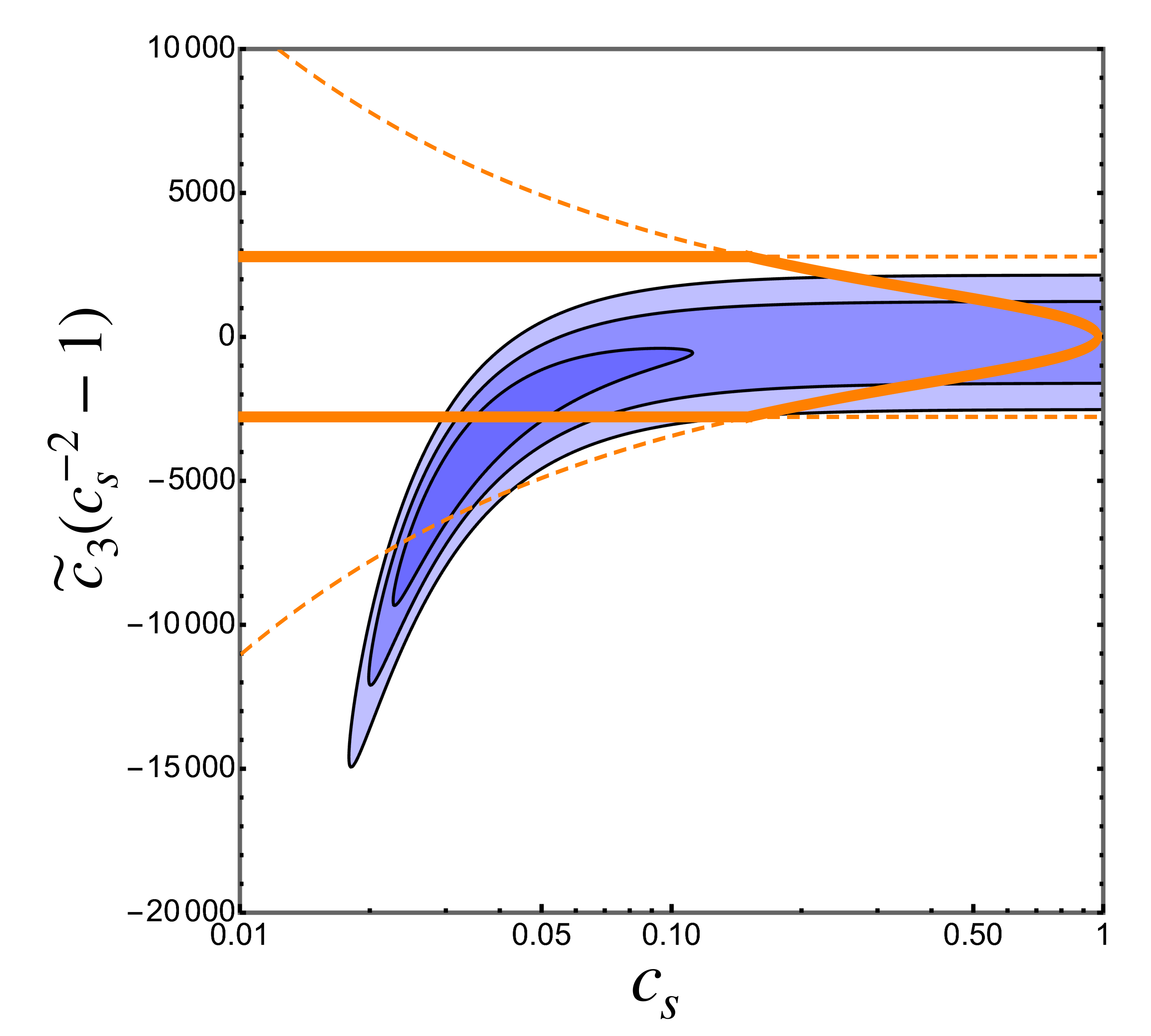}
	\caption{
The dark blue, blue, and light blue regions are the observationally allowed regions on the $c_s$-$\tilde{c}_3(c_s^{-2}-1)$ plane with the statistics of $1\sigma$, $2\sigma$, and $3\sigma$, respectively~\cite{Akrami:2019izv}.
The orange curve is a benchmark plot of the relation~\eqref{heavy_lambda3} for $m= 30 \,H$ and $\Lambda=1000\,H$.
The inside of the orange curve corresponds to $m=30\, H$ and $\Lambda>1000 \,H$. The cutoff scale is determined by $\Lambda_{\rm 6pt}$ for $c_s<c_s^\star\simeq 0.15$, whereas $\Lambda=\Lambda_{\rm 5pt}$ for $c_s>c_s^\star\simeq 0.15$.
}
	\label{eftnongaussianity}
\end{figure}

\medskip
To discuss implications to non-Gaussianities, let us rewrite the effective action as
\begin{align}
S_{\rm eff}&=\int dt d^3xa^3\Bigg[
-\frac{M_{\rm Pl}^2\dot{H}}{c_s^2}
\left(\dot{\pi}^2-c_s^2\frac{(\partial_i\pi)^2}{a^2}\right)
\nonumber
\\
&\qquad\qquad\qquad\quad
+M_{\rm Pl}^2\dot{H}(c_s^{-2}-1)
\left(\dot{\pi}\frac{(\partial_i\pi)^2}{a^2}
-\left(1+\frac{2}{3}\frac{\tilde{c}_3}{c_s^2}\right)\dot{\pi}^3
\right)+\mathcal{O}(\pi^4)
\Bigg]\,,
\end{align}
where the speed of sound $c_s$ and the dimensionless parameter $\tilde{c}_3$ are defined by
\begin{align}
c_s^{-2}&=1-\frac{2M_2^4}{M_{\rm Pl}^2\dot{H}}=1+4\frac{\dot{\theta}_0^2}{m^2}\,,
\\
\label{c_3}
\tilde{c}_3(c_s^{-2}-1)&=\frac{2c_s^2M_3^4}{M_{\rm Pl}^2\dot{H}}=-\frac{3}{4}\left(\lambda_3-\alpha\right)c_s^2(c_s^{-2}-1)^2\,.
\end{align}
In this language, the power spectrum is given by
\begin{align}
P_\zeta &= \frac{1}{c_s}\frac{H^4}{(2\pi)^2(-2M_{\rm Pl}^2\dot{H})}=\sqrt{1+4\pa{\frac{\dot\theta_0}{m}}^2}\frac{H^4}{(2\pi)^2r_0^2\dot\theta_0^2}\,.\label{heavy_powerspectrum_zeta}
\end{align} 
Also, the nonlinearity parameter $f_{\rm NL}$ reads
\begin{align}
f_{\rm NL}&=-\left(c_s^{-2}-1\right)\left[\frac{85}{324}+\frac{10}{243}\left(\tilde{c}_3+\frac{3}{2}c_s^2\right)\right]\,.
\end{align}
See Fig.~\ref{eftnongaussianity} for the current observational constraints on the parameters $(c_s,\tilde{c}_3)$~\cite{Akrami:2019izv}.

\paragraph{Scale of new physics.}

In the previous subsection, we demonstrated that the EFT with a nonzero $\tilde{c}_3$ cannot be UV completed by a single heavy scalar. Now let us discuss a relation between the value of $\tilde{c}_3$ and the scale of new physics beyond the UV two-field model. As in Eq.~\eqref{c_3}, $\tilde{c}_3$ is sourced by a cubic self-coupling $\lambda_3$ of the heavy isocurvature mode and a curvature $\alpha$ of the two-field space.

\medskip
First, let us consider the case when $\lambda_3\neq1$, but $\alpha=1$:
\begin{align}
|\wilde c_3(c_s^{-2} - 1)|
&= \frac{3}{4}|\lambda_3-1|(c_s^{-2} - 1)^2c_s^2 \,.
\end{align}
Then, it can be rephrased in terms of the cutoff scales~\eqref{cutoff_5pt}-\eqref{cutoff_lambda3_pre} as
\begin{align}
|\wilde c_3(c_s^{-2} - 1)|
&\simeq 3\times 10^{13}\cdot\frac{H^6(c_s^{-2}-1)^{1/2}c_s^{1/2}}{m^5\Lambda_{\rm 5pt}}\cdot\left(\frac{2\times 10^{-9}}{P_\zeta}\right)^{3/2}\,,
\\
\label{Lambda_c_3_lambda}
 |\wilde c_3(c_s^{-2} - 1)|
&\simeq  2\times 10^{18}\cdot\frac{H^8}{m^6\Lambda_{\rm 6pt}^2}\cdot\left(\frac{2\times 10^{-9}}{P_\zeta}\right)^{2}\,.
\end{align}
We find that six-point scattering provides a stronger constraint for small $c_s$, whereas five-point scattering gives a stronger one for $c_s\simeq1$. Then, the relation between $\tilde{c}_3$ and the cutoff scale $\Lambda$ is given by
\begin{align}
\label{heavy_lambda3}
|\wilde c_3(c_s^{-2} - 1)|
&
\simeq
\left\{\begin{array}{ll}
\displaystyle
3\times 10^{13}\cdot\frac{H^6(c_s^{-2}-1)^{1/2}c_s^{1/2}}{m^5\Lambda}\cdot\left(\frac{2\times 10^{-9}}{P_\zeta}\right)^{3/2}
&
{\rm for}\quad c_s>c_s^\star\,,
\\[5mm]
\displaystyle
2\times 10^{18}\cdot\frac{H^8}{m^6\Lambda^2}\cdot\left(\frac{2\times 10^{-9}}{P_\zeta}\right)^{2}
&
{\rm for}\quad c_s<c_s^\star\,,
\end{array}\right.
\end{align}
where the critical speed of sound $c_s^\star$ for a fixed cutoff scale $\Lambda$ is determined by\footnote{
The corresponding value $\tilde{c}_3^\star$ of $\tilde{c}_3$ is given by
\begin{align}
|{\wilde c_3}^{\,\star}({c_s^\star}^{-2}-1)|=4\times10^8\cdot\frac{H^4({c^\star_s}^{-2}-1)c^\star_s}{m^4} \cdot\left(\frac{2\times 10^{-9}}{P_\zeta}\right)\,.
\end{align}}
\begin{align}
({c_s^{\star}})^{-1}-c_s^\star
\simeq
6\times 10^9\cdot\frac{H^4}{m^2\Lambda^2}\cdot\frac{2\times 10^{-9}}{P_\zeta}\,.
\end{align}
For illustration, the relation~\eqref{heavy_lambda3} for $m=30H$ and $\Lambda=1000H$ is provided in Fig.~\ref{eftnongaussianity}.

\begin{figure}[t]
	\centering
	\includegraphics[width=9cm]{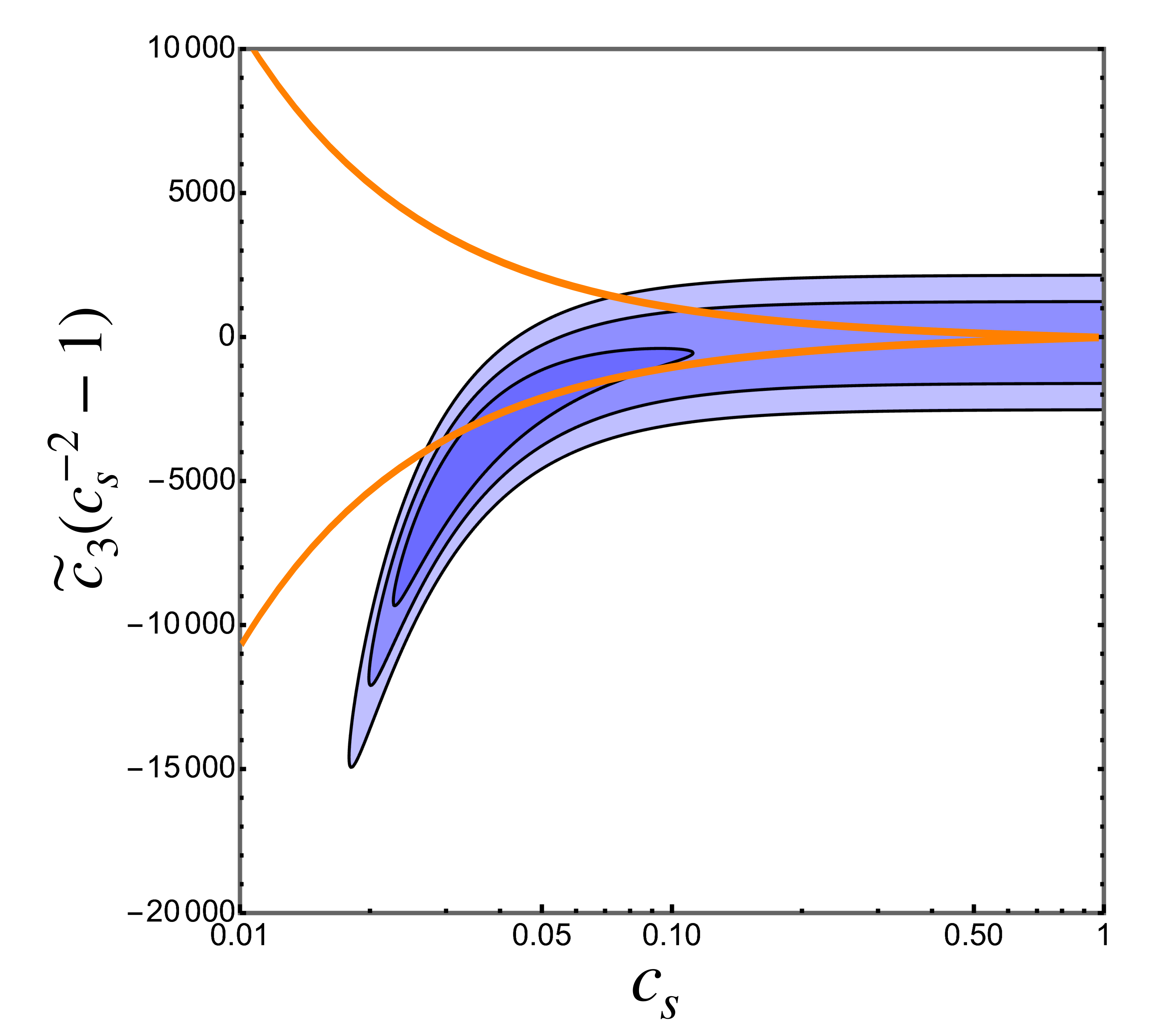}
	\caption{
The dark blue, blue, and light blue regions are the observationally allowed regions on the $c_s$-$\tilde{c}_3(c_s^{-2}-1)$ plane with the statistics of $1\sigma$, $2\sigma$, and $3\sigma$, respectively~\cite{Akrami:2019izv}.
The orange curve is a benchmark plot of the relation~\eqref{Lambda_c_3_alpha} for $m= 10 \,H$ and $\Lambda=300\,H$. The inside of the orange curve corresponds to $m=10\, H$ and $\Lambda>300 \,H$.
}
	\label{eftnongaussianity2}
\end{figure}

\medskip
Next we consider the case when $\alpha\neq1$, but $\lambda_3=1$. Then, the cutoff scale is dictated by four-point scattering as Eq.~\eqref{cutoff_curvature_pre}, which gives
\begin{align}
 |\wilde c_3(c_s^{-2} - 1)|
&= \frac{3}{4}|\alpha-1|(c_s^{-2} - 1)^2c_s^2
\nonumber
\\
\label{Lambda_c_3_alpha}
&\simeq 1\times 10^{9}\cdot\frac{H^4(c_s^{-2}-1)c_s}{m^2\Lambda_{\rm 4pt}^2}\cdot\left(\frac{2\times 10^{-9}}{P_\zeta}\right)\,.
\end{align}
We also find $c_s$-dependence which is analogous to $\dot{\theta}_0$ dependence of the relation~\eqref{f_NL_curvature}. The relation~\eqref{Lambda_c_3_alpha} for $m=10\,H$ and $\Lambda=300\,H$ is provided in Fig.~\ref{eftnongaussianity2}.

\paragraph{Perturbativity.}

Finally, we comment on the validity of the dictionary~\eqref{EFT_parameters}. In the heavy mass regime $m\gg H$, the equation of motion of the isocurvature $\sigma$ is approximated around the Hubble scale as
\begin{align}
\frac{\sigma}{r_0}&=-\frac{\dot{\theta}_0^2}{m^2}\left[
-2 \dot{\pi}-\dot{\pi}^2+\frac{(\partial_i\pi)^2}{a^2}
\right]
\nonumber
\\
&\quad
-\frac{3}{2}\lambda_3\left(\frac{\sigma}{r_0}\right)^2-\alpha\frac{\dot{\theta}_0^2}{m^2}\frac{\sigma}{r_0}\left[
-2 \dot{\pi}-\dot{\pi}^2+\frac{(\partial_i\pi)^2}{a^2}
\right]+\ldots\,,
\end{align}
where the dots stand for cubic and higher orders in perturbations. The dictionary~\eqref{EFT_parameters} was derived by assuming that the second line is smaller than the first line, which is justified if the following conditions are satisfied:
\begin{align}
\left|\frac{3}{2}\lambda_3\frac{\dot{\theta}_0^2}{m^2}\left[
-2 \dot{\pi}-\dot{\pi}^2+\frac{(\partial_i\pi)^2}{a^2}
\right]\right|\ll1\,,
\quad
\left|\alpha\frac{\dot{\theta}_0^2}{m^2}\left[
-2 \dot{\pi}-\dot{\pi}^2+\frac{(\partial_i\pi)^2}{a^2}
\right]\right|\ll1\,.
\end{align}
By noticing the linear order relation $\zeta\simeq-H\pi$ and the smallness of the scalar power spectrum $P_\zeta\simeq2\times10^{-9}$, these conditions imply
\begin{align}
\left|\tilde{c}_3(c_s^{-2}-1)\right|=3(1-c_s^2)\left|\left(\lambda_3-\alpha\right)\frac{\dot{\theta}_0^2}{m^2}\right|\ll \frac{1-c_s^2}{|\dot{\zeta}/H|}\,.
\end{align}
If we estimate the size of $\zeta$ as $\zeta\simeq P_\zeta^{1/2}$, the right hand side is $\mathcal{O}(10^4)$ in the regime of our interests. Therefore, the dictionary~\eqref{EFT_parameters} provides a good approximation at least within the regime allowed by the present observations.

\section{Summary and discussion}\label{prospects}

In this paper we studied implications of perturbative unitarity for two-scalar inflation models with the inflaton and one massive scalar. When the massive scalar has a Hubble scale mass $m\sim H$, the model is classified into quasi-single field inflation, which is of interests especially in the context of Cosmological Collider Program. While it has been expected that observable non-Gaussianities with $|f_{\rm NL}|\gtrsim1$ are sourced by the cubic self-coupling of the isocurvature mode, we have shown that non-Gaussianities sourced by the cubic coupling are too small to observe when the iscovurvature potential is renormalizable. In other words, observable non-Gaussianities with $|f_{\rm NL}|\gtrsim1$ require new physics at a higher energy scale. We determined the scale of new physics as Eq.~\eqref{cutoff_lambda3} based on perturbative unitarity.
Interestingly, we found that for $H\gtrsim 6\times 10^{9}$ GeV ($V^{1/4}\gtrsim 2\times 10^{14}$ GeV), Planck suppressed operators can easily generate too large non-Gaussanities and so it is hard to realize successful quasi-single field inflation without introducing a mechanism to suppress quantum gravity corrections. It would be interesting to explore such a UV mechanism explicitly.
On the other hand, for $H\lesssim 6\times 10^{8}$ GeV ($V^{1/4}\lesssim 5\times 10^{13}\,$GeV), observable non-Gaussianities $|f_{\rm NL}|\gtrsim 1$ (blow the present upper bound) can be generated by a new physics around the scale~\eqref{cutoff_lambda3}\footnote{
Indeed, it is not difficult to embed such an effective $\sigma^3$ interaction into a model with an additional heavy scalar $\Sigma$ with the mass $M\gg H$. For example, one can achieve this by introducing a linear mixing $\sigma\Sigma$ and a cubic self-coupling $\Sigma^3$ (to be precise, one needs to introduce a liner mixing through $\Sigma(2r_0\sigma+\sigma^2)$ to maintain renormalizablitiy). Note that in contrast to $\sigma^3$, the $\Sigma^3$ coupling is renormalizable by itself since $\Sigma$ has no kinetic mixing with $\theta$ and $\sigma$, so that there is no constraint on it at least if one allows fine-tuning.
}. Our result thus identifies the scale of the new physics that generates the effective cubic interaction $\sigma^3$ sourcing observable non-Gaussianities.
Besides, we also performed a similar analysis for non-Gaussianities sourced by the cubic coupling associated with the field space curvature, whose results are summarized in Eq.~\eqref{f_NL_curvature}.

\medskip
When the massive scalar is heavy $m\gg H$, the inflationary dynamics is captured by the effective field theory of the inflaton. Using perturbative unitarity, we identified the parameter space of the $P(X,\phi)$ model scanned by two-scalar UV models. In particular, the cubic and higher order terms in $X$ cannot be generated at IR unless we introduced new physics beyond the two-scalar UV model. The scale of such new physics beyond the two-scalar model can again be identified by perturbative unitarity and the results are summarized in Eqs.~\eqref{Lambda_c_3_lambda}-\eqref{Lambda_c_3_alpha}. 
It would be interesting to generalize our analysis to UV models with multiple scalars and clarify which parameter space of the $P(X,\phi)$ model can be UV completed by scalar field theories.
It would also be interesting to include heavy spinning fields in the discussion. By pushing this direction further, we would be able to enlarge the scope of the cosmological collider to higher energy.

\section*{Acknowledgements}
We would like to thank Xingang Chen for useful discussion and also for pointing out an error in the earlier draft.
S.K. is supported in part by the Senshu Scholarship Foundation. T.N. is supported in part by JSPS KAKENHI Grant Numbers JP17H02894 and 20H01902. S.Z. is supported in part by the Swedish Research Council under grants number 2015-05333 and 2018-03803.

\appendix
\section{Evaluation of cutoff scales}
\label{cutoffscale}
In this appendix, we derive the cutoff scales from unitarity of partial wave amplitudes taking care of numerical coefficients. For this purpose, it is convenient to use a canonically normalized basis of initial/final states which satisfy
\begin{align}
\langle \mathcal{A}|\mathcal{B}\rangle=(2\pi)^4\delta^4(p_{\mathcal{A}}-p_{\mathcal{B}})\delta_{\mathcal{A}\mathcal{B}}\,,
\end{align}
where $\mathcal{A},\mathcal{B}$ are discrete labels of the states. In such a basis, the unitarity bound for elastic scattering reads
\begin{align}
\label{unitarity_identical}
|M_{\mathcal{A}\mathcal{A}}|\le2\,,\quad 0\le {\rm Im } \,M_{\mathcal{A}\mathcal{A}}\le2\,,\quad -1\le {\rm Re } \,M_{\mathcal{A}\mathcal{A}}\le 1\,.
\end{align}
Similarly, the bound for $\mathcal{A}\neq\mathcal{B}$ reads
\begin{align}
\label{unitarity_general}
|M_{\mathcal{A}\mathcal{B}}|\leq 1\,.
\end{align}

\subsection{Four-point amplitude}

We first derive the cutoff scale from unitarity of four-point scattering amplitudes, whose partial wave expansion reads
\begin{align}
M(s,t)=16\pi\sum_{n=0}^\infty a_n\left(E_{\rm cm}^2\right)(2n+1)P_n(\cos \theta)\,.
\end{align}
Here $E_{\rm cm}$ is the total energy in the center-of-mass frame, $\theta$ is the scattering angle, and $P_n(x)$ is the Legendre polynomial. By employing the angular momentum eigenstates as a canonically normalized basis, the bound~\eqref{unitarity_general} for non-elastic scattering process is translated in terms of the partial wave amplitudes $a_n$ as
\begin{align}
\big|a_n(E_{\rm cm})\big|\le\frac{1}{2}\,,
\label{partial_unitarity_gene}
\end{align}
where we assumed $m\ll E_{\rm cm}$. For elastic scattering, we may use~\eqref{unitarity_identical} to derive the following bounds:
\begin{align}
\big|a_n(E_{\rm cm})\big| \le 1\,,\qquad 0\le {\rm Im } \,a_n(E_{\rm cm})\le1\,,\qquad -\frac{1}{2}\le {\rm Re } \,a_n(E_{\rm cm})\le \frac{1}{2}\,.
\label{partial_unitarity_same}
\end{align}
Now we apply these bounds to $2\to 2$ scattering given in Eq.~\eqref{4ptT2}. First, for $\sigma\sigma\to\varphi\varphi$ scattering, the amplitude is expanded as
\begin{align}
M_{\sigma\sigma\to\varphi\varphi} &=-(\alpha-1)\frac{E^2_{\rm cm}}{r_0^2}+ \mathcal O(E_{\rm cm}^0)\,,
\end{align}
which gives the partial wave amplitudes,
\begin{align}
a_0\left(E_{\rm cm}^2\right)=-\frac{(\alpha-1)E_{\rm cm}^2}{16\pi r_0^2}\,,\quad a_n\left(E_{\rm cm}^2\right)=0\quad (n\geq1)\,.
\end{align}
Then, Eq.~\eqref{partial_unitarity_gene} implies the following cutoff scale:
\begin{align}
\Lambda_{\sigma\sigma\to \varphi\varphi}=2\sqrt{2\pi}\cdot\frac{r_0}{|\alpha-1|^{1/2}}\,.
\end{align}
Similarly, for $\varphi\sigma\to\varphi\sigma$ scattering, the amplitude reads
\begin{align}
M_{\varphi\sigma\to\varphi\sigma}=\frac{(\alpha-1)E_{\rm cm}^2}{2r_0^2}-\frac{(\alpha-1)E_{\rm cm}^2}{2r_0^2}P_1(\cos \theta)+ \mathcal O(E_{\rm cm}^0)\,,
\end{align}
which gives
\begin{align}
a_0\left(E_{\rm cm}^2\right)=\frac{(\alpha-1)E_{\rm cm}^2}{32\pi r_0^2}\,,\quad a_1\left(E_{\rm cm}^2\right)=-\frac{(\alpha-1)E_{\rm cm}^2}{96\pi r_0^2}\,,\quad a_n\left(E_{\rm cm}^2\right)=0\quad (n\geq 2)\,.
\end{align}
Note that for this process the $S$-wave amplitude $a_0$ provides a stronger bound than $a_1$. Then, Eq.~\eqref{partial_unitarity_same} implies the following cutoff scale:
\begin{align}
\Lambda_{\varphi\sigma\to\varphi\sigma}=4\sqrt{\pi}\cdot\frac{r_0}{|\alpha-1|^{1/2}}\,.
\end{align}
We conclude that the cutoff scale dictated by four-point scattering is
\begin{align}
\Lambda_{\rm 4pt}=2\sqrt{2\pi}\cdot\frac{r_0}{|\alpha-1|^{1/2}}\,.
\end{align}

\subsection{Five-point and six-point amplitudes}

Let us perform a similar argument for five-point and six-point amplitudes. For simplicity, we focus on the $S$-wave amplitudes and introduce the following zero angular momentum states (see also Ref.~\cite{Chang:2019vez}):
\begin{align}
\ket{P;n_\varphi,n_\sigma}&=C_{n_\varphi,n_\sigma}\int d^4 xe^{iP.x}\Big[\varphi^{(-)}(x)\Big]^{n_\varphi}\Big[\sigma^{(-)}(x)\Big]^{n_\sigma}\ket{0}
\label{s-wavestate}\\
&=C_{n_\varphi,n_\sigma}\prod_i^{n_\varphi+n_\sigma}\bigg[\int\frac{d^3 \mathbf p_i}{2E(\mathbf p_i)(2\pi)^3}\bigg](2\pi)^4\delta^4\Big(P-\sum_i p_i\Big)\ket{p_1,\cdots,p_{n_\varphi+n_\sigma}}\,,
\nonumber
\end{align}
where the $(n_\varphi+n_\sigma)$-particle state $\ket{p_1,\cdots,p_{n_\varphi+n_\sigma}}$ is defined in the standard manner. Also $\varphi^{(-)}$ is the creation operator part of $\varphi$,
\begin{align}
\varphi^{(-)}(x)=\int\frac{d^3\mathbf p}{(2\pi)^3}\frac{1}{\sqrt{2E(\mathbf p)}}e^{-ip.x}a_{\varphi,p}^{\dagger}\,,
\end{align}
and similarly for $\sigma^{(-)}$.
$C_{n_\varphi,n_\sigma}$ is the normalization factor defined as
\begin{align}
\frac{1}{|C_{n_\varphi,n_\sigma}|^2}=n_\varphi!n_\sigma!\prod_i^{n_\varphi+n_\sigma}\bigg[\int\frac{d^3 \mathbf p_i}{2E(\mathbf p_i)(2\pi)^3}\bigg](2\pi)^4\delta^4(P-\sum_i p_i)
\end{align}
such that the states are canonically normalized:
\begin{align}
\braket{P^\prime;n^\prime_\varphi,n^\prime_\sigma|P;n_\varphi,n_\sigma}=(2\pi)^4\delta^4(P^\prime-P)\delta_{n_\varphi,n^\prime_\varphi}\delta_{n_\sigma,n^\prime_\sigma}\,.
\end{align}
More explicitly, $C_{n_\varphi,n_\sigma}$ for $n_\varphi+n_\sigma=2$ and $n_\varphi+n_\sigma=3$ are given by
\begin{align}
|C_{n_\varphi,n_\sigma}|^2&=\frac{8\pi}{n_\varphi! n_\sigma!}\qquad\qquad\qquad\,\,\text{for}\quad n_\varphi+n_\sigma=2\,,
\\
|C_{n_\varphi,n_\sigma}|^2&=\frac{16\pi}{n_\varphi! n_\sigma!}\left(\frac{4\pi}{E_{\rm cm}}\right)^2\qquad\text{for}\quad n_\varphi+n_\sigma=3\,.
\end{align}
In this language, the $S$-wave amplitudes are read off by projecting the initial/final states onto the zero angular momentum states~\eqref{s-wavestate}.

\paragraph{Five-point amplitudes.}
Now, we are ready to evaluate the cutoff scales from $S$-wave amplitudes. First, let us consider five-point scattering amplitudes. In the main text, we considered two amplitudes $M_{\sigma\sigma\sigma\varphi\varphi}$ and $M_{\sigma\varphi\varphi\varphi\varphi}$. While both of them break the unitarity bound at high energy for generic EFT parameters, high energy behavior of $M_{\sigma\sigma\sigma\varphi\varphi}$ depends not only on $\alpha$ and $\lambda_3$, but also on $\lambda_4$, which does not source bispectra. Therefore, it is not useful for exploring implications for the bispectrum $f_{\rm NL}$\footnote{
Besides, high energy behavior of $M_{\sigma\sigma\sigma\varphi\varphi}$ is dominated by the nonrenormalizable operator $\sigma^3(\partial_\mu\varphi)^2$ associated with the field space curvature, which we did not include in the main text. Note that one might wonder that $M_{\sigma\varphi\varphi\varphi\varphi}$ is also affected by a similar operator $\sigma\varphi^2(\partial_\mu\varphi)^2$, but it is prohibited by the shift symmetry of $\varphi$. Moreover, high energy behavior of this operator is as mild as the non-derivative interaction $\sigma\varphi^4$, essentially because these two are related to each other by partial integrals and field redefinition.}. On the other hand, $M_{\sigma\varphi\varphi\varphi\varphi}$ has no $\lambda_4$-dependence and so it is useful for the study of the bispectrum. Hence, we focus on $M_{\sigma\varphi\varphi\varphi\varphi}$ for the estimation of the cutoff scale in this paper. Also, in the regime of our interests, it is easy to see that the strongest unitarity constraint on $\alpha-1$ is obtained from four-point scattering, rather than five-point and six-point scattering, so that we set $\alpha=1$ and focus on the effect of $\lambda_3-1$ in the following discussion.

\medskip
Then, let us consider the $S$-wave amplitude associated with $M_{\sigma\varphi\varphi\varphi\varphi}$. First, let us consider $\sigma\varphi\to\varphi\varphi\varphi$ scattering. By using Eq.~\eqref{sig_phi4} and Eq.~\eqref{s-wavestate}, the $S$-wave component of the scattering amplitude $M_{\sigma\varphi\to\varphi\varphi\varphi}$ reads
\begin{align}
&M_{\sigma\varphi\to\varphi\varphi\varphi}^{(S)}
\nonumber
\\
&=C_{1,1}C_{3,0}\prod_{i=1}^{5}\bigg[\int\frac{d^3 \mathbf p_i}{2E(\mathbf p_i)(2\pi)^3}\bigg]
(2\pi)^4\delta^4(p_1+p_2-P)
(2\pi)^4\delta^4(p_3+p_4+p_5+P)M_{\sigma\varphi\to\varphi\varphi\varphi}
\nonumber\\
&= \frac{9}{64\sqrt{3}\pi^2}\frac{m^2\pa{1-\lambda_3}}{r_0^3}E_{\rm cm} + \mathcal O(E_{\rm cm}^0)\,.
\label{plane_to_partial1}
\end{align}
Similarly, the $S$-wave component of $M_{\sigma\varphi\varphi\to\varphi\varphi}$ reads
\begin{align}
M_{\sigma\varphi\varphi\to\varphi\varphi}^{(S)}&= -\frac{9}{64\sqrt{2}\pi^2}\frac{m^2\pa{\lambda_3-1}}{r_0^3}E_{\rm cm} + \mathcal O(E_{\rm cm}^0)\,.
\label{plane_to_partial2}
\end{align}
Then, the unitarity bounds $|M_{\sigma\varphi\to\varphi\varphi\varphi}^{(S)}|\leq1$ and $|M_{\sigma\varphi\varphi\to\varphi\varphi}^{(S)}|\leq 1$ imply the cutoff scale as
\begin{align}
\Lambda_{\rm 5pt}&=\frac{64\sqrt{2}\pi^2}{9}\cdot\frac{ r_0^3}{m^2|\lambda_3-1|}
\,.
\end{align}

\paragraph{Six-point scattering.}

Finally, let us look at six-point amplitudes, among which the relevant one for our purpose is $M_{\varphi\varphi\varphi\varphi\varphi\varphi}$ given in Eq.~\eqref{phi6}. The $S$-wave component for $\varphi\varphi\varphi\to\varphi\varphi\varphi$ scattering is
\begin{align}
M_{\varphi\varphi\varphi\to\varphi\varphi\varphi}^{(S)}&=-\frac{15}{512\pi^3}\frac{m^2(\lambda_3-1)}{r_0^4}E_{\rm cm}^2+\mathcal{O}(E_{\rm cm}^{0})\,.
\label{plane_to_partial4}
\end{align}
Then, the unitarity bound $|{\rm Re}\,M_{\varphi\varphi\varphi\to\varphi\varphi\varphi\varphi}^{(S)}|\leq1$ implies the cutoff scale as
\begin{align}
\Lambda_{\rm 6pt}&=\frac{16\sqrt{2}\pi^{3/2}}{\sqrt{15}}\cdot\frac{r_0^2}{m|\lambda_3-1|^{1/2}}\,.
\end{align}

\section{Field redefinition for amplitude computation}
\label{app:redefinition}

In this appendix we summarize field redefinition useful for the computation of scattering amplitudes. Essentially because we are interested in high-energy scattering $E\gg m,|\dot{\theta}_0|$, one may explicitly show that terms which contain $\dot{\theta}_0$ are irrelevant for evaluation of the cutoff scale. Then, it is enough to analyze high-energy scattering in the following setup:
\begin{align}
\label{Lorentz}
S&=\int dt d^3xa^3\left[
-\frac12 \left(1+\frac{\sigma}{r_0}\right)^2\del_\mu\varphi\del^\mu\varphi - \frac12  \del_\mu\sigma\del^\mu\sigma
-\frac{\alpha-1}{2}\frac{\sigma^2}{r_0^2}(\partial_\mu\varphi)^2
 - V(\sigma)
\right]\,.
\end{align} 
To simplify the computation of scattering amplitudes, we perform the field redefinition,
\begin{align}
\label{redef}
(r_0+\sigma)e^{i\varphi/r_0}=r_0+\varphi_1+i\varphi_2\,,
\end{align}
or in other words,
\begin{align}
\sigma=\sqrt{(r_0+\varphi_1)^2+\varphi_2^2}-r_0\,,
\quad
\varphi=r_0\arctan \frac{\varphi_2}{r_0+\varphi_1}\,.
\end{align}
Note that at the linear level we have $\sigma\simeq\varphi_1$ and $\varphi\simeq\varphi_2$, so that scattering amplitudes of $\sigma$ and $\varphi$ are identical to those of $\varphi_1$ and $\varphi_2$.

\paragraph{Derivative interactions.}

We begin by the first two terms in Eq.~\eqref{Lorentz}, which can be reformulated by the field redefinition~\eqref{redef} as
\begin{align}
\label{flat_redef}
-\frac12 \left(1+\frac{\sigma}{r_0}\right)^2\del_\mu\varphi\del^\mu\varphi - \frac12  \del_\mu\sigma\del^\mu\sigma
=-\frac{1}{2}(\partial\varphi_1)^2-\frac{1}{2}(\partial\varphi_2)^2
\,.
\end{align}
Note that it is simply a canonical kinetic term  without any interaction. On the other hand, the third term in Eq.~\eqref{Lorentz} reads
\begin{align}
&-\frac{\alpha-1}{2}\frac{\sigma^2}{r_0^2}(\partial_\mu\varphi)^2
\nonumber
\\
&=
-\frac{\alpha-1}{2}
\frac{(\sqrt{(r_0+\varphi_1)^2+\varphi_2^2}-r_0)^2}{r_0^2}\frac{\Big(\partial_\mu\frac{r_0\varphi_2}{r_0+\varphi_1}\Big)^2}{\left(1+\frac{\varphi_2^2}{(r_0+\varphi_1)^2}\right)^2}
\nonumber
\\
&=-\frac{\alpha-1}{2}\left[\frac{\varphi_1^2}{r_0^2}(\partial_\mu\varphi_2)^2
+\frac{-2\varphi_1^3+\varphi_1\varphi_2^2}{r_0^3}(\partial_\mu\varphi_2)^2
-\frac{2\varphi_1^2\varphi_2}{r_0^3}(\partial_\mu\varphi_1\partial^\mu\varphi_2)
\right.
\nonumber
\\
&\quad
\left.+\frac{3\varphi_1^4-5\varphi_1^2\varphi_2^2+\tfrac{1}{4}\varphi_2^4}{r_0^4}(\partial_\mu\varphi_2)^2
+\frac{6\varphi_1^3\varphi_2-2\varphi_1\varphi_2^3}{r_0^4}(\partial_\mu\varphi_1\partial^\mu\varphi_2)
+\frac{\varphi_1^2\varphi_2^2}{r_0^4}(\partial_\mu\varphi_1)^2
\right]
\nonumber
\\
\label{curved_redef}
&\quad
+\mathcal{O}(\varphi_i^7)
\,.
\end{align}
Therefore, derivative interactions appear only when the field space curvature is nonzero. In this expression, e.g., the $\mathcal{O}(E^2) $ contribution to the four-point amplitude~\eqref{4ptT2} is sourced only by the four-point contact vertex presented in the first term of the third line. Similar simplification occurs when we evaluate more general amplitudes too.

\paragraph{Isocurvature potential.}

Next, let us take a look at the isocurvature potential,
\begin{align}
V(\sigma)=V\left(\sqrt{(r_0+\varphi_1)^2+\varphi_2^2}-r_0\right)
\end{align}
with a parameterization
\begin{align}
V(\sigma)=\frac{1}{2}m^2r_0^2\left[
\frac{\sigma^2}{r_0^2}+\lambda_3\frac{\sigma^3}{r_0^3}+\frac{\lambda_4}{4}\frac{\sigma^4}{r_0^4}
\right]\,.
\end{align}
More explicitly, we find
\begin{align}
V(\sigma)&=
\frac{1}{2}m^2r_0^2
\left[
\frac{\varphi_1^2}{r_0^2}
+\frac{\varphi_1^3+\varphi_1\varphi_2^2}{r_0^3}
+\frac{\varphi_1^4+2\varphi_1^2\varphi_2^2+\varphi_2^4}{4r_0^4}
\right]
\nonumber
\\
&\quad
+\frac{\lambda_3-1}{2}m^2r_0^2
\left[\frac{\varphi_1^3}{r_0^3}+\frac{3}{2}\frac{\varphi_1^2\varphi_2^2}{r_0^4}
+\frac{3}{4}\frac{-2\varphi_1^3\varphi_2^2+\varphi_1\varphi_2^4}{r_0^5}
+\frac{1}{8}\frac{12\varphi_1^4\varphi_2^2-15\varphi_1^2\varphi_2^4+\varphi_2^6}{r_0^6}\right]
\nonumber
\\
\label{potential_redef}
&\quad
+\frac{\lambda_4-1}{8}m^2r_0^2
\left[
\frac{\varphi_1^4}{r_0^4}
+2\frac{\varphi_1^3\varphi_2^2}{r_0^5}
+\frac{1}{2}\frac{-4\varphi_1^4\varphi_2^2+3\varphi_1^2\varphi_2^4}{r_0^6}\right]
+\mathcal{O}(\varphi_i^7)\,,
\end{align}
from which it is obvious in particular that non-renormalizable interactions do not appear when $\alpha=\lambda_3=\lambda_4=1$ is satisfied.

\medskip
In the main text, we provided Feynman diagrams before the field redefinition for illustration. However, the expressions~\eqref{flat_redef}, \eqref{curved_redef}, and~\eqref{potential_redef} simplify the computation of scattering amplitudes a lot (the results are of course invariant under field redefinition).

\bibliography{qsi}{}

\providecommand{\href}[2]{#2}\begingroup\raggedright\begin{thebibliography}{100}

\bibitem{Lee:1977eg}
B.~W. Lee, C.~Quigg and H.~B. Thacker,  {\em {Weak Interactions at Very
  High-Energies: The Role of the Higgs Boson Mass}}, Phys. Rev. {\bf D16}
  (1977)
1519.

\bibitem{Lee:1977yc}
B.~W. Lee, C.~Quigg and H.~B. Thacker,  {\em {The Strength of Weak Interactions
  at Very High-Energies and the Higgs Boson Mass}}, Phys. Rev. Lett. {\bf 38}
  (1977)
883--885.

\bibitem{Dicus:1992vj}
D.~A. Dicus and V.~S. Mathur,  {\em {Upper bounds on the values of masses in
  unified gauge theories}}, Phys. Rev. {\bf D7} (1973)
3111--3114.

\bibitem{Chanowitz:1985hj}
M.~S. Chanowitz and M.~K. Gaillard,  {\em {The TeV Physics of Strongly
  Interacting W's and Z's}}, Nucl. Phys. {\bf B261} (1985)
379--431.

\bibitem{Chatrchyan:2013lba}
{CMS} Collaboration, S.~Chatrchyan {\em et al.},  {\em {Observation of a New
  Boson with Mass Near 125 GeV in $pp$ Collisions at $\sqrt{s}$ = 7 and 8
  TeV}}, JHEP {\bf 06} (2013) 081
  [\href{http://www.arXiv.org/abs/1303.4571}{{\tt 1303.4571}}].

\bibitem{Chatrchyan:2012ufa}
{CMS} Collaboration, S.~Chatrchyan {\em et al.},  {\em {Observation of a New
  Boson at a Mass of 125 GeV with the CMS Experiment at the LHC}}, Phys. Lett.
  B {\bf 716} (2012) 30--61 [\href{http://www.arXiv.org/abs/1207.7235}{{\tt
  1207.7235}}].

\bibitem{Aad:2012tfa}
{ATLAS} Collaboration, G.~Aad {\em et al.},  {\em {Observation of a new
  particle in the search for the Standard Model Higgs boson with the ATLAS
  detector at the LHC}}, Phys. Lett. B {\bf 716} (2012) 1--29
  [\href{http://www.arXiv.org/abs/1207.7214}{{\tt 1207.7214}}].

\bibitem{Maldacena:2002vr}
J.~M. Maldacena,  {\em {Non-Gaussian features of primordial fluctuations in
  single field inflationary models}}, JHEP {\bf 05} (2003) 013
  [\href{http://www.arXiv.org/abs/astro-ph/0210603}{{\tt astro-ph/0210603}}].

\bibitem{Lerner:2010mq}
R.~N. Lerner and J.~McDonald,  {\em {A Unitarity-Conserving Higgs Inflation
  Model}}, Phys. Rev. D {\bf 82} (2010) 103525
  [\href{http://www.arXiv.org/abs/1005.2978}{{\tt 1005.2978}}].

\bibitem{Giudice:2010ka}
G.~F. Giudice and H.~M. Lee,  {\em {Unitarizing Higgs Inflation}}, Phys. Lett.
  B {\bf 694} (2011) 294--300 [\href{http://www.arXiv.org/abs/1010.1417}{{\tt
  1010.1417}}].

\bibitem{Atkins:2010yg}
M.~Atkins and X.~Calmet,  {\em {Remarks on Higgs Inflation}}, Phys. Lett. B
  {\bf 697} (2011) 37--40 [\href{http://www.arXiv.org/abs/1011.4179}{{\tt
  1011.4179}}].

\bibitem{Calmet:2013hia}
X.~Calmet and R.~Casadio,  {\em {Self-healing of unitarity in Higgs
  inflation}}, Phys. Lett. B {\bf 734} (2014) 17--20
  [\href{http://www.arXiv.org/abs/1310.7410}{{\tt 1310.7410}}].

\bibitem{Barbon:2015fla}
J.~L.~F. Barbon, J.~A. Casas, J.~Elias-Miro and J.~R. Espinosa,  {\em {Higgs
  Inflation as a Mirage}}, JHEP {\bf 09} (2015) 027
  [\href{http://www.arXiv.org/abs/1501.02231}{{\tt 1501.02231}}].

\bibitem{Fumagalli:2017cdo}
J.~Fumagalli, S.~Mooij and M.~Postma,  {\em {Unitarity and predictiveness in
  new Higgs inflation}}, JHEP {\bf 03} (2018) 038
  [\href{http://www.arXiv.org/abs/1711.08761}{{\tt 1711.08761}}].

\bibitem{Lee:2018esk}
H.~M. Lee,  {\em {Light inflaton completing Higgs inflation}}, Phys. Rev. D
  {\bf 98} (2018), no.~1, 015020
  [\href{http://www.arXiv.org/abs/1802.06174}{{\tt 1802.06174}}].

\bibitem{Ema:2020zvg}
Y.~Ema, K.~Mukaida and J.~van~de Vis,  {\em {Higgs inflation as nonlinear sigma
  model and scalaron as its $\sigma$-meson}}, JHEP {\bf 11} (2020) 011
  [\href{http://www.arXiv.org/abs/2002.11739}{{\tt 2002.11739}}].

\bibitem{Weinberg:1979sa}
S.~Weinberg,  {\em {Baryon and Lepton Nonconserving Processes}}, Phys. Rev.
  Lett. {\bf 43} (1979) 1566--1570.

\bibitem{Buchmuller:1985jz}
W.~Buchmuller and D.~Wyler,  {\em {Effective Lagrangian Analysis of New
  Interactions and Flavor Conservation}}, Nucl. Phys. B {\bf 268} (1986)
  621--653.

\bibitem{Grzadkowski:2010es}
B.~Grzadkowski, M.~Iskrzynski, M.~Misiak and J.~Rosiek,  {\em {Dimension-Six
  Terms in the Standard Model Lagrangian}}, JHEP {\bf 10} (2010) 085
  [\href{http://www.arXiv.org/abs/1008.4884}{{\tt 1008.4884}}].

\bibitem{Grinstein:2007iv}
B.~Grinstein and M.~Trott,  {\em {A Higgs-Higgs bound state due to new physics
  at a TeV}}, Phys. Rev. D {\bf 76} (2007) 073002
  [\href{http://www.arXiv.org/abs/0704.1505}{{\tt 0704.1505}}].

\bibitem{deFlorian:2016spz}
{LHC Higgs Cross Section Working Group} Collaboration, D.~de~Florian {\em et
  al.},  {\em {Handbook of LHC Higgs Cross Sections: 4. Deciphering the Nature
  of the Higgs Sector}}, \href{http://www.arXiv.org/abs/1610.07922}{{\tt
  1610.07922}}.

\bibitem{Brivio:2017vri}
I.~Brivio and M.~Trott,  {\em {The Standard Model as an Effective Field
  Theory}}, Phys. Rept. {\bf 793} (2019) 1--98
  [\href{http://www.arXiv.org/abs/1706.08945}{{\tt 1706.08945}}].

\bibitem{Chang:2019vez}
S.~Chang and M.~A. Luty,  {\em {The Higgs Trilinear Coupling and the Scale of
  New Physics}},
\href{http://www.arXiv.org/abs/1902.05556}{{\tt 1902.05556}}.

\bibitem{Nagai:2019tgi}
R.~Nagai, M.~Tanabashi, K.~Tsumura and Y.~Uchida,  {\em {Symmetry and geometry
  in a generalized Higgs effective field theory: Finiteness of oblique
  corrections versus perturbative unitarity}}, Phys. Rev. {\bf D100} (2019),
  no.~7, 075020
[\href{http://www.arXiv.org/abs/1904.07618}{{\tt 1904.07618}}].

\bibitem{Chen:2009zp}
X.~Chen and Y.~Wang,  {\em {Quasi-Single Field Inflation and
  Non-Gaussianities}}, JCAP {\bf 1004} (2010) 027
[\href{http://www.arXiv.org/abs/0911.3380}{{\tt 0911.3380}}].

\bibitem{Baumann:2011nk}
D.~Baumann and D.~Green,  {\em {Signatures of Supersymmetry from the Early
  Universe}}, Phys. Rev. {\bf D85} (2012) 103520
[\href{http://www.arXiv.org/abs/1109.0292}{{\tt 1109.0292}}].

\bibitem{Noumi:2012vr}
T.~Noumi, M.~Yamaguchi and D.~Yokoyama,  {\em {Effective field theory approach
  to quasi-single field inflation and effects of heavy fields}}, JHEP {\bf 06}
  (2013) 051
[\href{http://www.arXiv.org/abs/1211.1624}{{\tt 1211.1624}}].

\bibitem{Arkani-Hamed:2015bza}
N.~Arkani-Hamed and J.~Maldacena,  {\em {Cosmological Collider Physics}},
\href{http://www.arXiv.org/abs/1503.08043}{{\tt 1503.08043}}.

\bibitem{Chen:2009we}
X.~Chen and Y.~Wang,  {\em {Large non-Gaussianities with Intermediate Shapes
  from Quasi-Single Field Inflation}}, Phys. Rev. {\bf D81} (2010) 063511
[\href{http://www.arXiv.org/abs/0909.0496}{{\tt 0909.0496}}].

\bibitem{Assassi:2012zq}
V.~Assassi, D.~Baumann and D.~Green,  {\em {On Soft Limits of Inflationary
  Correlation Functions}}, JCAP {\bf 1211} (2012) 047
[\href{http://www.arXiv.org/abs/1204.4207}{{\tt 1204.4207}}].

\bibitem{Sefusatti:2012ye}
E.~Sefusatti, J.~R. Fergusson, X.~Chen and E.~P.~S. Shellard,  {\em {Effects
  and Detectability of Quasi-Single Field Inflation in the Large-Scale
  Structure and Cosmic Microwave Background}}, JCAP {\bf 1208} (2012) 033
[\href{http://www.arXiv.org/abs/1204.6318}{{\tt 1204.6318}}].

\bibitem{Norena:2012yi}
J.~Norena, L.~Verde, G.~Barenboim and C.~Bosch,  {\em {Prospects for
  constraining the shape of non-Gaussianity with the scale-dependent bias}},
  JCAP {\bf 1208} (2012) 019
[\href{http://www.arXiv.org/abs/1204.6324}{{\tt 1204.6324}}].

\bibitem{Emami:2013lma}
R.~Emami,  {\em {Spectroscopy of Masses and Couplings during Inflation}}, JCAP
  {\bf 1404} (2014) 031
[\href{http://www.arXiv.org/abs/1311.0184}{{\tt 1311.0184}}].

\bibitem{Liu:2015tza}
J.~Liu, Y.~Wang and S.~Zhou,  {\em {Inflation with Massive Vector Fields}},
  JCAP {\bf 1508} (2015) 033
[\href{http://www.arXiv.org/abs/1502.05138}{{\tt 1502.05138}}].

\bibitem{Dimastrogiovanni:2015pla}
E.~Dimastrogiovanni, M.~Fasiello and M.~Kamionkowski,  {\em {Imprints of
  Massive Primordial Fields on Large-Scale Structure}}, JCAP {\bf 1602} (2016)
  017
[\href{http://www.arXiv.org/abs/1504.05993}{{\tt 1504.05993}}].

\bibitem{Schmidt:2015xka}
F.~Schmidt, N.~E. Chisari and C.~Dvorkin,  {\em {Imprint of inflation on galaxy
  shape correlations}}, JCAP {\bf 1510} (2015), no.~10, 032
[\href{http://www.arXiv.org/abs/1506.02671}{{\tt 1506.02671}}].

\bibitem{Chen:2015lza}
X.~Chen, M.~H. Namjoo and Y.~Wang,  {\em {Quantum Primordial Standard Clocks}},
  JCAP {\bf 1602} (2016), no.~02, 013
[\href{http://www.arXiv.org/abs/1509.03930}{{\tt 1509.03930}}].

\bibitem{Bonga:2015urq}
B.~Bonga, S.~Brahma, A.-S. Deutsch and S.~Shandera,  {\em {Cosmic variance in
  inflation with two light scalars}}, JCAP {\bf 1605} (2016), no.~05, 018
[\href{http://www.arXiv.org/abs/1512.05365}{{\tt 1512.05365}}].

\bibitem{Delacretaz:2015edn}
L.~V. Delacretaz, T.~Noumi and L.~Senatore,  {\em {Boost Breaking in the EFT of
  Inflation}}, JCAP {\bf 1702} (2017), no.~02, 034
[\href{http://www.arXiv.org/abs/1512.04100}{{\tt 1512.04100}}].

\bibitem{Flauger:2016idt}
R.~Flauger, M.~Mirbabayi, L.~Senatore and E.~Silverstein,  {\em {Productive
  Interactions: heavy particles and non-Gaussianity}}, JCAP {\bf 1710} (2017),
  no.~10, 058
[\href{http://www.arXiv.org/abs/1606.00513}{{\tt 1606.00513}}].

\bibitem{Lee:2016vti}
H.~Lee, D.~Baumann and G.~L. Pimentel,  {\em {Non-Gaussianity as a Particle
  Detector}}, JHEP {\bf 12} (2016) 040
[\href{http://www.arXiv.org/abs/1607.03735}{{\tt 1607.03735}}].

\bibitem{Delacretaz:2016nhw}
L.~V. Delacretaz, V.~Gorbenko and L.~Senatore,  {\em {The Supersymmetric
  Effective Field Theory of Inflation}}, JHEP {\bf 03} (2017) 063
[\href{http://www.arXiv.org/abs/1610.04227}{{\tt 1610.04227}}].

\bibitem{Meerburg:2016zdz}
P.~D. Meerburg, M.~Münchmeyer, J.~B. Muñoz and X.~Chen,  {\em {Prospects for
  Cosmological Collider Physics}}, JCAP {\bf 1703} (2017), no.~03, 050
[\href{http://www.arXiv.org/abs/1610.06559}{{\tt 1610.06559}}].

\bibitem{Chen:2016uwp}
X.~Chen, Y.~Wang and Z.-Z. Xianyu,  {\em {Standard Model Background of the
  Cosmological Collider}}, Phys. Rev. Lett. {\bf 118} (2017), no.~26, 261302
[\href{http://www.arXiv.org/abs/1610.06597}{{\tt 1610.06597}}].

\bibitem{Chen:2016hrz}
X.~Chen, Y.~Wang and Z.-Z. Xianyu,  {\em {Standard Model Mass Spectrum in
  Inflationary Universe}}, JHEP {\bf 04} (2017) 058
[\href{http://www.arXiv.org/abs/1612.08122}{{\tt 1612.08122}}].

\bibitem{An:2017hlx}
H.~An, M.~McAneny, A.~K. Ridgway and M.~B. Wise,  {\em {Quasi Single Field
  Inflation in the non-perturbative regime}}, JHEP {\bf 06} (2018) 105
[\href{http://www.arXiv.org/abs/1706.09971}{{\tt 1706.09971}}].

\bibitem{Tong:2017iat}
X.~Tong, Y.~Wang and S.~Zhou,  {\em {On the Effective Field Theory for
  Quasi-Single Field Inflation}}, JCAP {\bf 1711} (2017), no.~11, 045
[\href{http://www.arXiv.org/abs/1708.01709}{{\tt 1708.01709}}].

\bibitem{Iyer:2017qzw}
A.~V. Iyer, S.~Pi, Y.~Wang, Z.~Wang and S.~Zhou,  {\em {Strongly Coupled
  Quasi-Single Field Inflation}}, JCAP {\bf 1801} (2018), no.~01, 041
[\href{http://www.arXiv.org/abs/1710.03054}{{\tt 1710.03054}}].

\bibitem{An:2017rwo}
H.~An, M.~McAneny, A.~K. Ridgway and M.~B. Wise,  {\em {Non-Gaussian
  Enhancements of Galactic Halo Correlations in Quasi-Single Field Inflation}},
  Phys. Rev. {\bf D97} (2018), no.~12, 123528
[\href{http://www.arXiv.org/abs/1711.02667}{{\tt 1711.02667}}].

\bibitem{Kumar:2017ecc}
S.~Kumar and R.~Sundrum,  {\em {Heavy-Lifting of Gauge Theories By Cosmic
  Inflation}}, JHEP {\bf 05} (2018) 011
[\href{http://www.arXiv.org/abs/1711.03988}{{\tt 1711.03988}}].

\bibitem{Riquelme:2017bxt}
S.~Riquelme~M.,  {\em {Non-Gaussianities in a two-field generalization of
  Natural Inflation}}, JCAP {\bf 1804} (2018), no.~04, 027
[\href{http://www.arXiv.org/abs/1711.08549}{{\tt 1711.08549}}].

\bibitem{Saito:2018omt}
R.~Saito and T.~Kubota,  {\em {Heavy Particle Signatures in Cosmological
  Correlation Functions with Tensor Modes}}, JCAP {\bf 1806} (2018), no.~06,
  009
[\href{http://www.arXiv.org/abs/1804.06974}{{\tt 1804.06974}}].

\bibitem{Cabass:2018roz}
G.~Cabass, E.~Pajer and F.~Schmidt,  {\em {Imprints of Oscillatory Bispectra on
  Galaxy Clustering}}, JCAP {\bf 1809} (2018), no.~09, 003
[\href{http://www.arXiv.org/abs/1804.07295}{{\tt 1804.07295}}].

\bibitem{Dimastrogiovanni:2018uqy}
E.~Dimastrogiovanni, M.~Fasiello and G.~Tasinato,  {\em {Probing the
  inflationary particle content: extra spin-2 field}}, JCAP {\bf 1808} (2018),
  no.~08, 016
[\href{http://www.arXiv.org/abs/1806.00850}{{\tt 1806.00850}}].

\bibitem{Bordin:2018pca}
L.~Bordin, P.~Creminelli, A.~Khmelnitsky and L.~Senatore,  {\em {Light
  Particles with Spin in Inflation}}, JCAP {\bf 1810} (2018), no.~10, 013
[\href{http://www.arXiv.org/abs/1806.10587}{{\tt 1806.10587}}].

\bibitem{Arkani-Hamed:2018kmz}
N.~Arkani-Hamed, D.~Baumann, H.~Lee and G.~L. Pimentel,  {\em {The Cosmological
  Bootstrap: Inflationary Correlators from Symmetries and Singularities}},
\href{http://www.arXiv.org/abs/1811.00024}{{\tt 1811.00024}}.

\bibitem{Kumar:2018jxz}
S.~Kumar and R.~Sundrum,  {\em {Seeing Higher-Dimensional Grand Unification In
  Primordial Non-Gaussianities}},
\href{http://www.arXiv.org/abs/1811.11200}{{\tt 1811.11200}}.

\bibitem{Goon:2018fyu}
G.~Goon, K.~Hinterbichler, A.~Joyce and M.~Trodden,  {\em {Shapes of gravity:
  Tensor non-Gaussianity and massive spin-2 fields}},
\href{http://www.arXiv.org/abs/1812.07571}{{\tt 1812.07571}}.

\bibitem{Wu:2018lmx}
Y.-P. Wu,  {\em {Higgs as heavy-lifted physics during inflation}},
\href{http://www.arXiv.org/abs/1812.10654}{{\tt 1812.10654}}.

\bibitem{Chua:2018dqh}
W.~Z. Chua, Q.~Ding, Y.~Wang and S.~Zhou,  {\em {Imprints of Schwinger Effect
  on Primordial Spectra}},
\href{http://www.arXiv.org/abs/1810.09815}{{\tt 1810.09815}}.

\bibitem{Wang:2018tbf}
Y.~Wang, Y.-P. Wu, J.~Yokoyama and S.~Zhou,  {\em {Hybrid Quasi-Single Field
  Inflation}}, JCAP {\bf 1807} (2018), no.~07, 068
[\href{http://www.arXiv.org/abs/1804.07541}{{\tt 1804.07541}}].

\bibitem{McAneny:2019epy}
M.~McAneny and A.~K. Ridgway,  {\em {New Shapes of Primordial Non-Gaussianity
  from Quasi-Single Field Inflation with Multiple Isocurvatons}},
\href{http://www.arXiv.org/abs/1903.11607}{{\tt 1903.11607}}.

\bibitem{Li:2019ves}
L.~Li, T.~Nakama, C.~M. Sou, Y.~Wang and S.~Zhou,  {\em {Gravitational
  Production of Superheavy Dark Matter and Associated Cosmological
  Signatures}},
\href{http://www.arXiv.org/abs/1903.08842}{{\tt 1903.08842}}.

\bibitem{Kim:2019wjo}
S.~Kim, T.~Noumi, K.~Takeuchi and S.~Zhou,  {\em {Heavy Spinning Particles from
  Signs of Primordial Non-Gaussianities: Beyond the Positivity Bounds}},
\href{http://www.arXiv.org/abs/1906.11840}{{\tt 1906.11840}}.

\bibitem{Sleight:2019mgd}
C.~Sleight,  {\em {A Mellin Space Approach to Cosmological Correlators}},
\href{http://www.arXiv.org/abs/1906.12302}{{\tt 1906.12302}}.

\bibitem{Biagetti:2019bnp}
M.~Biagetti,  {\em {The Hunt for Primordial Interactions in the Large Scale
  Structures of the Universe}}, Galaxies {\bf 7} (2019), no.~3, 71
[\href{http://www.arXiv.org/abs/1906.12244}{{\tt 1906.12244}}].

\bibitem{Sleight:2019hfp}
C.~Sleight and M.~Taronna,  {\em {Bootstrapping Inflationary Correlators in
  Mellin Space}},
\href{http://www.arXiv.org/abs/1907.01143}{{\tt 1907.01143}}.

\bibitem{Welling:2019bib}
Y.~Welling,  {\em {A simple, exact, model of quasi-single field inflation}},
\href{http://www.arXiv.org/abs/1907.02951}{{\tt 1907.02951}}.

\bibitem{Alexander:2019vtb}
S.~Alexander, S.~J. Gates, L.~Jenks, K.~Koutrolikos and E.~McDonough,  {\em
  {Higher Spin Supersymmetry at the Cosmological Collider: Sculpting SUSY
  Rilles in the CMB}}, JHEP {\bf 10} (2019) 156
[\href{http://www.arXiv.org/abs/1907.05829}{{\tt 1907.05829}}].

\bibitem{Lu:2019tjj}
S.~Lu, Y.~Wang and Z.-Z. Xianyu,  {\em {A Cosmological Higgs Collider}}, JHEP
  {\bf 02} (2020) 011
[\href{http://www.arXiv.org/abs/1907.07390}{{\tt 1907.07390}}].

\bibitem{Hook:2019zxa}
A.~Hook, J.~Huang and D.~Racco,  {\em {Searches for other vacua II: A new
  Higgstory at the cosmological collider}},
\href{http://www.arXiv.org/abs/1907.10624}{{\tt 1907.10624}}.

\bibitem{Hook:2019vcn}
A.~Hook, J.~Huang and D.~Racco,  {\em {Minimal signatures of the Standard Model
  in non-Gaussianities}},
\href{http://www.arXiv.org/abs/1908.00019}{{\tt 1908.00019}}.

\bibitem{ScheihingHitschfeld:2019tzr}
B.~Scheihing~Hitschfeld, {\em {Revealing the Structure of the Inflationary
  Landscape through Primordial non-Gaussianity}}.
\newblock PhD thesis, Chile U., Santiago, 2019.
\newblock
\href{http://www.arXiv.org/abs/1909.11223}{{\tt 1909.11223}}.
\newblock

\bibitem{Baumann:2019oyu}
D.~Baumann, C.~D. Pueyo, A.~Joyce, H.~Lee and G.~L. Pimentel,  {\em {The
  Cosmological Bootstrap: Weight-Shifting Operators and Scalar Seeds}},
\href{http://www.arXiv.org/abs/1910.14051}{{\tt 1910.14051}}.

\bibitem{Wang:2019gbi}
L.-T. Wang and Z.-Z. Xianyu,  {\em {In Search of Large Signals at the
  Cosmological Collider}}, JHEP {\bf 02} (2020) 044
[\href{http://www.arXiv.org/abs/1910.12876}{{\tt 1910.12876}}].

\bibitem{Liu:2019fag}
T.~Liu, X.~Tong, Y.~Wang and Z.-Z. Xianyu,  {\em {Probing P and CP Violations
  on the Cosmological Collider}},
\href{http://www.arXiv.org/abs/1909.01819}{{\tt 1909.01819}}.

\bibitem{Wang:2019gok}
D.-G. Wang,  {\em {On the inflationary massive field with a curved field
  manifold}}, JCAP {\bf 2001} (2020), no.~01, 046
[\href{http://www.arXiv.org/abs/1911.04459}{{\tt 1911.04459}}].

\bibitem{Wang:2020uic}
Y.~Wang and Y.~Zhu,  {\em {Cosmological Collider Signatures of Massive Vectors
  from Non-Gaussian Gravitational Waves}},
\href{http://www.arXiv.org/abs/2001.03879}{{\tt 2001.03879}}.

\bibitem{Li:2020xwr}
L.~Li, S.~Lu, Y.~Wang and S.~Zhou,  {\em {Cosmological Signatures of Superheavy
  Dark Matter}},
\href{http://www.arXiv.org/abs/2002.01131}{{\tt 2002.01131}}.

\bibitem{Baumann:2020dch}
D.~Baumann, C.~Duaso~Pueyo, A.~Joyce, H.~Lee and G.~L. Pimentel,  {\em {The
  Cosmological Bootstrap: Spinning Correlators from Symmetries and
  Factorization}}, \href{http://www.arXiv.org/abs/2005.04234}{{\tt
  2005.04234}}.

\bibitem{Kogai:2020vzz}
K.~Kogai, K.~Akitsu, F.~Schmidt and Y.~Urakawa,  {\em {Galaxy imaging surveys
  as spin-sensitive detector for cosmological colliders}},
  \href{http://www.arXiv.org/abs/2009.05517}{{\tt 2009.05517}}.

\bibitem{Aoki:2020zbj}
S.~Aoki and M.~Yamaguchi,  {\em {Disentangling mass spectra of multiple fields
  in cosmological collider}}, \href{http://www.arXiv.org/abs/2012.13667}{{\tt
  2012.13667}}.

\bibitem{Maru:2021ezc}
N.~Maru and A.~Okawa,  {\em {Non-Gaussianity from $X, Y$ gauge bosons in
  Cosmological Collider Physics}},
  \href{http://www.arXiv.org/abs/2101.10634}{{\tt 2101.10634}}.

\bibitem{Chen:2006nt}
X.~Chen, M.-x. Huang, S.~Kachru and G.~Shiu,  {\em {Observational signatures
  and non-Gaussianities of general single field inflation}}, JCAP {\bf 01}
  (2007) 002 [\href{http://www.arXiv.org/abs/hep-th/0605045}{{\tt
  hep-th/0605045}}].

\bibitem{Brennan:2017rbf}
T.~D. Brennan, F.~Carta and C.~Vafa,  {\em {The String Landscape, the
  Swampland, and the Missing Corner}}, PoS {\bf TASI2017} (2017) 015
  [\href{http://www.arXiv.org/abs/1711.00864}{{\tt 1711.00864}}].

\bibitem{Palti:2019pca}
E.~Palti,  {\em {The Swampland: Introduction and Review}}, Fortsch. Phys. {\bf
  67} (2019), no.~6, 1900037 [\href{http://www.arXiv.org/abs/1903.06239}{{\tt
  1903.06239}}].

\bibitem{Baumann:2011su}
D.~Baumann and D.~Green,  {\em {Equilateral Non-Gaussianity and New Physics on
  the Horizon}}, JCAP {\bf 1109} (2011) 014
[\href{http://www.arXiv.org/abs/1102.5343}{{\tt 1102.5343}}].

\bibitem{Baumann:2014cja}
D.~Baumann, D.~Green and R.~A. Porto,  {\em {B-modes and the Nature of
  Inflation}}, JCAP {\bf 01} (2015) 016
  [\href{http://www.arXiv.org/abs/1407.2621}{{\tt 1407.2621}}].

\bibitem{Koehn:2015vvy}
M.~Koehn, J.-L. Lehners and B.~Ovrut,  {\em {Nonsingular bouncing cosmology:
  Consistency of the effective description}}, Phys. Rev. D {\bf 93} (2016),
  no.~10, 103501 [\href{http://www.arXiv.org/abs/1512.03807}{{\tt
  1512.03807}}].

\bibitem{Baumann:2015nta}
D.~Baumann, D.~Green, H.~Lee and R.~A. Porto,  {\em {Signs of Analyticity in
  Single-Field Inflation}}, Phys. Rev. {\bf D93} (2016), no.~2, 023523
[\href{http://www.arXiv.org/abs/1502.07304}{{\tt 1502.07304}}].

\bibitem{deRham:2017aoj}
C.~de~Rham and S.~Melville,  {\em {Unitary null energy condition violation in
  P(X) cosmologies}}, Phys. Rev. D {\bf 95} (2017), no.~12, 123523
  [\href{http://www.arXiv.org/abs/1703.00025}{{\tt 1703.00025}}].

\bibitem{Fumagalli:2020ody}
J.~Fumagalli, M.~Postma and M.~Van Den~Bout,  {\em {Matching and running
  sensitivity in non-renormalizable inflationary models}}, JHEP {\bf 09} (2020)
  114 [\href{http://www.arXiv.org/abs/2005.05905}{{\tt 2005.05905}}].

\bibitem{Chen:2012ge}
X.~Chen and Y.~Wang,  {\em {Quasi-Single Field Inflation with Large Mass}},
  JCAP {\bf 1209} (2012) 021
[\href{http://www.arXiv.org/abs/1205.0160}{{\tt 1205.0160}}].

\bibitem{Pi:2012gf}
S.~Pi and M.~Sasaki,  {\em {Curvature Perturbation Spectrum in Two-field
  Inflation with a Turning Trajectory}}, JCAP {\bf 1210} (2012) 051
[\href{http://www.arXiv.org/abs/1205.0161}{{\tt 1205.0161}}].

\bibitem{Lyth:1996im}
D.~H. Lyth,  {\em {What would we learn by detecting a gravitational wave signal
  in the cosmic microwave background anisotropy?}}, Phys. Rev. Lett. {\bf 78}
  (1997) 1861--1863 [\href{http://www.arXiv.org/abs/hep-ph/9606387}{{\tt
  hep-ph/9606387}}].

\bibitem{Tristram:2020wbi}
M.~Tristram {\em et al.},  {\em {Planck constraints on the tensor-to-scalar
  ratio}}, \href{http://www.arXiv.org/abs/2010.01139}{{\tt 2010.01139}}.

\bibitem{Cheung:2007st}
C.~Cheung, P.~Creminelli, A.~L. Fitzpatrick, J.~Kaplan and L.~Senatore,  {\em
  {The Effective Field Theory of Inflation}}, JHEP {\bf 03} (2008) 014
  [\href{http://www.arXiv.org/abs/0709.0293}{{\tt 0709.0293}}].

\bibitem{Tolley:2009fg}
A.~J. Tolley and M.~Wyman,  {\em {The Gelaton Scenario: Equilateral
  non-Gaussianity from multi-field dynamics}}, Phys. Rev. {\bf D81} (2010)
  043502
[\href{http://www.arXiv.org/abs/0910.1853}{{\tt 0910.1853}}].

\bibitem{Gong:2013sma}
J.-O. Gong, S.~Pi and M.~Sasaki,  {\em {Equilateral non-Gaussianity from heavy
  fields}}, JCAP {\bf 1311} (2013) 043
[\href{http://www.arXiv.org/abs/1306.3691}{{\tt 1306.3691}}].

\bibitem{Akrami:2019izv}
{Planck} Collaboration, Y.~Akrami {\em et al.},  {\em {Planck 2018 results. IX.
  Constraints on primordial non-Gaussianity}},
\href{http://www.arXiv.org/abs/1905.05697}{{\tt 1905.05697}}.

\end{thebibliography}\endgroup
\bibliographystyle{utphys}

\end{document}